\documentclass[iop]{emulateapj}

\usepackage[colorlinks,urlcolor=blue,citecolor=blue,linkcolor=blue]{hyperref} 
\usepackage{natbib,aas_macros,verbatim,url}
\usepackage{times,color,amsmath}
\usepackage{verbatim,subfigure}
\usepackage[printonlyused,nohyperlinks]{acronym}

\newcounter{aff}
\newcommand{\rightascension}{19:06:48.86(4)}
\newcommand{\declination}{07:46:25.9(7)}
\newcommand{\period}{0.14407315538(5)}
\newcommand{\freq}{6.940918295(2)}
\newcommand{\fo}{$-$9.7643(9) }
\newcommand{\foo}{5.0(7)      }
\newcommand{\fooo}{$-$1.8(3)  }
\newcommand{\foooo}{$-$1.7(2) }
\newcommand{\dm}{217.7508(4)}
\newcommand{\epoch}{54289.000001}
\newcommand{\orbitalperiod}{0.16599304683(11)}
\newcommand{\semimajoraxis}{1.4199620(18)}
\newcommand{\eccentricity}{0.0853028(6)}
\newcommand{\epochperiastron}{54288.9298810(2)}
\newcommand{\periastron}{76.3320(6)}
\newcommand{\omdot}{7.5841(5)}
\newcommand{\pbdot}{$-$0.56(3)}
\newcommand{\einstein}{0.000470(5)}
\newcommand{\massfunction}{0.1115662(2)}

\newcommand{\mpulsar}{}

\newcommand{\characteristicage}{112.6}
\newcommand{\surfB}{1.73}
\newcommand{\dtopulsar}{$5.40^{+0.56}_{-0.60}$}
\newcommand{\dtopulsarnoerr}{5.40}

\newcommand{\grrightascension}{19:06:48.86(4)}
\newcommand{\grdeclination}{07:46:25.9(7)}
\newcommand{\grperiod}{0.14407315538(5)}
\newcommand{\grfreq}{6.940918295(2)}
\newcommand{\grfo}{$-$9.7642(9) }
\newcommand{\grfoo}{4.9(7)      }
\newcommand{\grfooo}{$-$1.8(3)  }
\newcommand{\grfoooo}{$-$1.7(2) }
\newcommand{\grdm}{217.7508(4)}
\newcommand{\grepoch}{54289.000001}
\newcommand{\grorbitalperiod}{0.16599304686(11)}
\newcommand{\grsemimajoraxis}{1.4199506(18)}
\newcommand{\greccentricity}{0.0852996(6)}
\newcommand{\grepochperiastron}{54288.9298808(2)}
\newcommand{\grperiastron}{76.3317(6)}
\newcommand{\grmtotal}{2.6134(3)}
\newcommand{\grmcompanion}{1.322(11)}
\newcommand{\grmpulsar}{1.291(11)}
\newcommand{\grmassfunction}{0.1115636(4)}
\newcommand{\grmpulsartable}{$1.291^{+0.011}_{-0.011}$}
\newcommand{\grmcompaniontable}{$1.322^{+0.011}_{-0.011}$}
\newcommand{\grpbdot}{$-$0.56498(15)}
\newcommand{\grxpbdot}{0.03(3)}
\newcommand{\gromdot}{7.5844(5)}
\newcommand{\greinstein}{0.000470(5)}
\newcommand{\grinclination}{43.7(5)}
\newcommand{\grcharacteristicage}{112.6}
\newcommand{\grsurfB}{1.73}
\newcommand{\grdtopulsar}{$5.40^{+0.56}_{-0.60}$}
\newcommand{\HIdtopulsar}{$7.4^{+2.5}_{-1.4}$}

\newcommand{\J}{\mbox{J1906+0746}}

\def\be{\begin{equation}}
\def\ee{\end{equation}}
\def\ba{\begin{eqnarray}}
\def\ea{\end{eqnarray}}

\newcommand{\msun}{\ifmmode\mbox{M}_{\odot}\else$\mbox{M}_{\odot}$\fi}
\newcommand{\rsun}{\ifmmode\mbox{R}_{\odot}\else$\mbox{M}_{\odot}$\fi}
\newcommand{\degrees}{\ifmmode^{\circ}\else$^{\circ}$\fi}
\newcommand{\amin}{\ifmmode^{\prime}\else$^{\prime}$\fi}
\newcommand{\asec}{\ifmmode^{\prime\prime}\else$^{\prime\prime}$\fi}

\begin{document}

\shorttitle{The Young, Relativistic Binary Pulsar J1906+0746}
\shortauthors{van Leeuwen et al.}
\title{The Binary Companion of Young, Relativistic Pulsar J1906+0746}

%% Use refs to institutes, like on LPPS paper
\author{
J.~van~Leeuwen\altaffilmark{\ref{astron},\ref{uva}}, 
L.~Kasian\altaffilmark{\ref{ubc}}, 
I.~H.~Stairs\altaffilmark{\ref{ubc}}, 
D.~R.~Lorimer\altaffilmark{\ref{wvu}},
F.~Camilo\altaffilmark{\ref{ao},\ref{cny}},
S.~Chatterjee\altaffilmark{\ref{cornell}},
I.~Cognard\altaffilmark{   \ref{nancay}},
G.~Desvignes\altaffilmark{ \ref{mpifr}},
P.~C.~C.~Freire\altaffilmark{\ref{mpifr}},
G.~H.~Janssen\altaffilmark{\ref{astron}, \ref{jbo}}, 
M.~Kramer\altaffilmark{    \ref{mpifr}},
A.~G.~Lyne\altaffilmark{   \ref{jbo}}, 
D.~J.~Nice\altaffilmark{   \ref{lc}},
S.~M.~Ransom\altaffilmark{  \ref{cv}},
B.~W.~Stappers\altaffilmark{\ref{jbo}},
and
J.~M.~Weisberg\altaffilmark{\ref{clt}}
}

% first one is different?
\altaffiltext{1}{\refstepcounter{aff}\label{astron}\refstepcounter{aff}ASTRON, the Netherlands Institute for Radio Astronomy, Postbus 2, 7990 AA, Dwingeloo, The Netherlands; \href{mailto:leeuwen@astron.nl}{leeuwen@astron.nl}} 
\altaffiltext{\theaff}{\label{uva}\refstepcounter{aff}Astronomical Institute ``Anton Pannekoek'', University of Amsterdam, Science Park 904, 1098 XH Amsterdam, The Netherlands}
\altaffiltext{\theaff}{\label{ubc}\refstepcounter{aff}Department~of Physics and Astronomy, University~of British Columbia, Vancouver, BC V6T 1Z1, Canada} 
\altaffiltext{\theaff}{\label{wvu}\refstepcounter{aff}Department of Physics, West Virginia University, Morgantown, WV 26506, USA} 
\altaffiltext{\theaff}{\label{ao}\refstepcounter{aff}Arecibo Observatory, HC3 Box 53995, Arecibo, PR 00612, USA}
\altaffiltext{\theaff}{\label{cny}\refstepcounter{aff}Columbia Astrophysics Laboratory, Columbia University, New York, NY 10027, USA} 
\altaffiltext{\theaff}{\label{cornell}\refstepcounter{aff}Center for Radiophysics and Space Research, Cornell University, Ithaca, NY 14853, USA}\altaffiltext{\theaff}{\label{nancay}\refstepcounter{aff}Laboratoire de Physique et Chimie de l'Environnement et de l'Espace LPC2E CNRS-Universit\'e d'Orl\'eans, 45071 Orl\'eans, and Station de radioastronomie de Nan\c cay, Observatoire de Paris, CNRS/INSU, 18330 Nan\c cay, France}
\altaffiltext{\theaff}{\label{mpifr}\refstepcounter{aff}Max-Planck-Institut f\"ur Radioastronomie, D-53121 Bonn, Germany} 
\altaffiltext{\theaff}{\label{jbo}\refstepcounter{aff}Jodrell Bank Centre for Astrophysics, School of Physics and Astronomy, University of Manchester, Manchester, M13 9PL, UK} 
\altaffiltext{\theaff}{\label{lc}\refstepcounter{aff}Department of Physics, Lafayette College, Easton, PA 18042, USA}
\altaffiltext{\theaff}{\label{cv}\refstepcounter{aff}NRAO (National Radio Astronomy Observatory), Charlottesville, VA 22903, USA} 
\altaffiltext{\theaff}{\label{clt}\refstepcounter{aff}Department of
  Physics and Astronomy, Carleton College, Northfield, MN 55057, USA}

\setcounter{aff}{0}
\setcounter{footnote}{0}

\begin{abstract}

PSR~\J\ is a young pulsar in the relativistic
binary with the {second-shortest known orbital period}, of 3.98 hours. We here
present a timing study based on five years of
observations, conducted with the 5 largest radio telescopes in the
world, aimed at determining the companion nature.  Through the
measurement of three post-Keplerian orbital parameters we find
the pulsar mass to be \grmpulsar\,$M_\odot$, and the companion mass
\grmcompanion\,$M_\odot$ respectively.  These masses 
fit well in the observed collection of \aclp{DNS},
but are also compatible
with other white dwarfs around young pulsars such as \J .
Neither radio
pulsations nor dispersion-inducing outflows that could have
further established the companion nature were detected. We derive an HI-absorption
distance, which indicates {that} an optical confirmation of a white dwarf
companion is very challenging. The pulsar is fading fast due to
geodetic precession, limiting future timing improvements.  We conclude
that young pulsar J1906+0746 is likely part of a {double} neutron star,
or is otherwise orbited by an older white dwarf, in an exotic system formed
through two stages of mass transfer.

\end{abstract}


\section{Introduction}
\label{sec:1906:introduction}

\newacro{ALFA}[ALFA]{Arecibo L-band Feed Array}
\newacro{ASKAP}[ASKAP]{Australian Square Kilometre Array Pathfinder}
\newacro{ASP}[ASP]{Arecibo Signal Processor}
\newacro{ATNF}[ATNF]{Australia Telescope National Facility}
\newacro{AXP}[AXP]{anomalous X-ray pulsar}
\newacro{BON}[BON]{Berkeley-Orl\'{e}ans-Nan\c{c}ay}
\newacro{CE}[CE]{common envelope}
\newacro{DM}[DM]{dispersion measure}
\newacro{DNS}[DNS]{double neutron star}
\newacro{EC}[EC]{electron-capture}
\newacro{FFT}[FFT]{fast Fourier transform}
\newacro{FT}[FT]{Fourier transform}
\newacro{GASP}[GASP]{Green Bank Astronomical Signal Processor}
\newacro{GBT}[GBT]{Green Bank Telescope}
\newacro{GC}[GC]{globular cluster}
\newacro{GR}[GR]{general relativity}
\newacro{IMBP}[IMBP]{intermediate-mass binary pulsar}
\newacro{ISM}[ISM]{interstellar medium}
\newacro{LMBP}[LMBP]{low-mass binary pulsar}
\newacro{LMXB}[LMXB]{low-mass X-ray binary}
\newacro{LOFAR}[LOFAR]{Low Frequency Array}
\newacro{MeerKAT}[MeerKAT]{Karoo Array Telescope}
\newacro{MJD}[MJD]{modified Julian day}
\newacro{MSP}[MSP]{millisecond pulsar}
\newacro{NRAO}[NRAO]{National Radio Astronomy Observatory}
\newacro{NS}[NS]{neutron star}
\newacro{PALFA}[PALFA]{Pulsar Arecibo L-band Feed Array}
\newacro{PMPS}[PMPS]{Parkes Multibeam Pulsar Survey}
\newacro{PDF}[PDF]{probability density function}
\newacro{RFI}[RFI]{radiofrequency interference}
\newacro{RMS}[RMS]{root mean square}
\newacro{RRAT}[RRAT]{rotating radio transient}
\newacro{RVM}[RVM]{rotating vector model}
\newacro{SEP}[SEP]{Strong Equivalence Principle}
\newacro{SGR}[SGR]{soft gamma repeater}
\newacro{SKA}[SKA]{Square Kilometre Array}
\newacro{SNR}[SNR]{signal-to-noise ratio}
\newacro{SSB}[SSB]{Solar System Barycentre}
\newacro{TOA}[TOA]{time of arrival}
\newacro{VLBI}[VLBI]{Very Long Baseline Interferometry}
\newacro{WAPP}[WAPP]{Wideband Arecibo Pulsar Processor}
\newacro{WD}[WD]{white dwarf}

Binaries harboring a neutron star are
windows on dynamical star systems
that have undergone and survived at least one supernova.  
By precisely measuring pulsar times of
arrival (\acsp{TOA}), and fitting binary models to these,
one can describe the orbital motions of the
pulsars and their companions -- and hence constrain their
masses.  In combination with
other information such as the pulsar spin, orbit, and companion nature,
these mass estimates can elucidate the binary interaction and mass
transfer history. 

In the vast majority of observed binary pulsar systems, the pulsar is
the first-born compact object: there, it is found to have been spun
up by accretion from its companion to a higher 
spin rate than seen in young
pulsars. These spun-up pulsars have far lower magnetic fields than the
general pulsar population. They thus show very stable rotation and
evolve only very slowly -- resulting in higher characteristic ages of
$\sim$10\,Gyr. This stability and longevity means 
such {\it recycled} pulsars are observable for much longer periods of
time than the high-field fast-evolving {\it young} pulsars.

The amount of recycling is related to the binary type: \acf{LMBP}
systems host millisecond pulsars (\acsp{MSP}), which have spin periods
of about 1$-$10 ms, and are orbited by a low-mass \acfp{WD}.  Pulsars
with more massive \ac{WD} companions or \acf{NS} companions, generally
have longer spin periods (10$-$200\,ms), and lower characteristic ages
\citep{lor08}.  This is consistent with a spin-up picture in which the
amount of mass transferred from the companion to the pulsar depends
largely on the duration of the mass-transfer stage, and hence, on the
mass of the {progenitor of the} companion (\citealt{acrs82}).

The 144-ms pulsar J1906+0746, discovered in precursor PALFA
observations in 2004 \citep{cfl+06,lcl+06,lsf+06}, is one of only a
handful of known {double-degenerate} relativistic binaries where the
pulsar is believed to be the younger of the two compact objects.
Other such pulsars, listed in Table \ref{table:binaries}, include the
\acl{WD} binaries J1141$-$6545 and B2303+46, and the second pulsar in
the double pulsar system, J0737$-$3039B.  Pulsar \J\ is in a 3.98-hour
binary orbit with an eccentricity of 0.085, making it the relativistic
binary pulsar with the second-shortest known orbital period -- second
only {to} the 2.4-h orbit double pulsar J0737$-$3039 (Table
\ref{table:binaries}).

\begin{table*}
\centering
\begin{tabular}{lcccccc}
\hline
\hline
{\bf Pulsar} &  {\bf Period} & {\bf $P_b$ } & {\bf Eccentricity} & {\bf Pulsar} & {\bf Companion} & {\bf Companion}\\ 
      & {\bf (ms)} & {\bf (days)} &  & {\bf Mass ($M_\odot$)} & {\bf Mass ($M_\odot$)} & {\bf Type}\\
\hline
\multicolumn{7}{c}{{\bf Young Pulsars in Relativistic Binaries }}\\
J0737$-$3039B$^1$ &  2773.5 & 0.102  &  0.088    &  $1.2489^{+0.0007}_{-0.0007}$  &  $1.3381^{+0.0007}_{-0.0007}$ & NS \\[1ex]
J1906+0746$^2$    &  144.1  & 0.166  &  0.085    &  \grmpulsartable  & \grmcompaniontable & WD or NS\\[1ex]
J1141$-$6545$^3$  &  393.9  & 0.198  &  0.17     &  $1.27^{+0.01}_{-0.01}$  & $1.02^{+0.01}_{0.01}$ & WD\\[1ex]
B2303+46$^{4}$    &  1066.4  & 12.3   &  0.66     & $1.34^{+0.10}_{-0.10}$  & $1.3^{+0.10}_{-0.10}$ & WD\\
\hline
\multicolumn{7}{c}{{\bf Recycled Pulsars in Relativistic Double Neutron Star Binaries}}\\
J0737$-$3039A$^1$ &  22.7  & 0.102  &  0.088    &  $1.3381^{+0.0007}_{-0.0007}$  & $1.2489^{+0.0007}_{-0.0007}$  & NS\\[1ex]
J1756$-$2251$^5$ &  28.5 & 0.320    &  0.18     &  $1.341^{+0.007}_{-0.007}$ & $1.230^{+0.007}_{-0.007}$  & NS\\[1ex]
B1913+16$^{6}$ &  59.0   & 0.323     &  0.62     & $1.439^{+0.0002}_{-0.0002}$ & $1.3886^{+0.0002}_{-0.0002}$ & NS \\[1ex]
B2127+11C$^{7}$ & 30.5   & 0.335      &  0.68    & $1.358^{+0.010}_{-0.010}$ & $1.354^{0.010}_{0.010}$ & NS\\
B1534+12$^8$ &  37.9    & 0.421        &  0.27     &  $1.3330^{+0.0002}_{-0.0002}$  & $1.3455^{+0.0002}_{-0.0002}$ & NS\\[1ex]
\hline
\multicolumn{7}{c}{{\bf Recycled Pulsars in Long-Period ($P_b > $1 day) Double Neutron Star Binaries}}\\
J1518+4904$^9$ &  40.9      & 8.63    &  0.25     & $0.72^{+0.51}_{-0.58}$ & $2.00^{+0.58}_{-0.51}$ & NS\\
 \multicolumn{4}{c}{} &  \multicolumn{2}{c}{ {\bf Total Mass ($M_\odot$)}} & \\[1ex]
J1829+2456$^{10}$   &  41.0  & 1.18    &  0.14     & \multicolumn{2}{c}{$2.5^{+0.2}_{-0.2}$} & NS\\
J1753$-$2240$^{11}$ &  95.1 & 13.6     & 0.30    &  \multicolumn{2}{c}{Not measured} & NS\\[1ex]
J1811$-$1736$^{12}$ &  104.2  & 18.8   &  0.82     & \multicolumn{2}{c}{$2.57^{+0.10}_{-0.10}$} & NS\\[1ex]
\hline
    \end{tabular}
 \caption[Known pulsars in relativistic and/or double neutron star
   binary systems]{Known pulsars in relativistic and/or double neutron
   star binary systems, ordered by pulsar age, with minor ordering on
   binary period. (Values from: 
$^1$\citealt{ksm+06};
   $^{2}$this work; 
   $^3$\citealt{bbv08}; 
   $^{4}$\citealt{tamt93},  \citealt{kk98}; 
   $^5$\citealt{2014MNRAS.443.2183F};
   $^{6}$\citealt{wnt10}; 
   $^7$\citealt{jcj+06}; 
   $^8$  \citealt{2014ApJ...787...82F}; 
   $^9$\citealt{jsk+08};
   $^{10}$\citealt{clm+05};
   $^{11}$\citealt{kkl+09}; 
   $^{12}$\citealt{cks+07}%
%   $^{11}$\citealt{ckl+04a};
)
 }
\label{table:binaries}
 \end{table*}

\begin{figure}[tb]
   \centering
   \vspace{10mm}
   \includegraphics[width=0.32\textwidth]{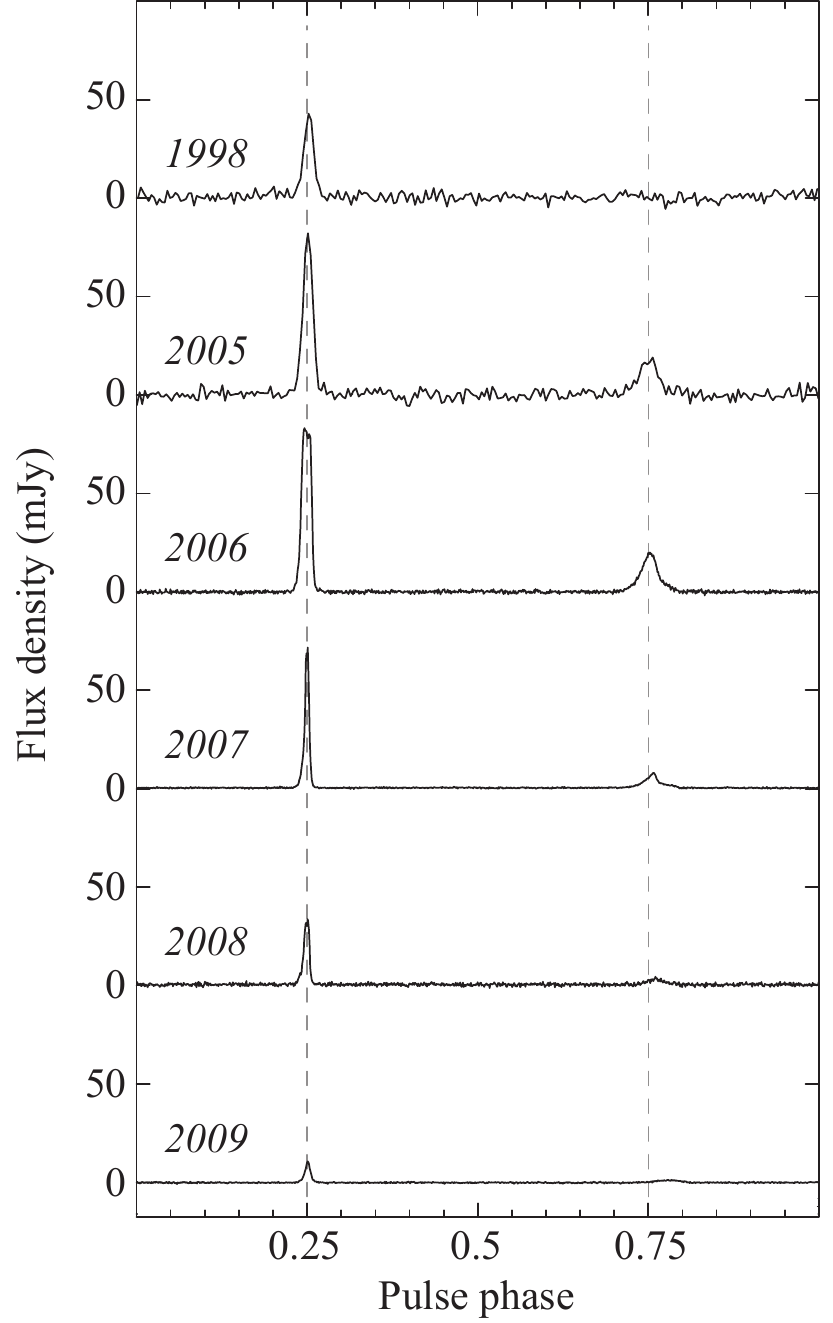}
   \caption{The profile evolution of \J. Shown are the 1998 and 2005
     Parkes profiles \citep{lsf+06}, and yearly Arecibo profiles (this
     work). All subplots are the same scale. The 1998 and 2005
     profiles used a different recording setup, with
     somewhat higher noise.}
   \label{fig:profile_changes}
\end{figure}

Data taken in 1998 and 2005 at Parkes showed rapid profile
evolution \citep[][and Fig.~\ref{fig:profile_changes}]{lsf+06}.
As
there was no sign of mode changing behaviour in the dense sampling
after the discovery, 
and thus no sign of magnetospheric instabilities \citep[e.g.][]{lhk+10},
these profile changes indicated a change in the
pulsar-observer viewing geometry.  Such \mbox{years-timescale} profile
changes are likely due to geodetic precession, the general
relativistic effect that causes spinning objects to precess about the
total angular momentum vector of the system \citep{dr74}, an effect
seen in other binary pulsars (J0737$-$3039B, \citealt{bkk+08}; J1141$-$6545, \citealt{hbo05};
 B1534+12, \citealt{sta04}; 
and B1913+16, \citealt{wrt89,kra98}).

Given its high spin-down rate ($\dot{P} \sim 2\times 10^{-14}$ s/s)
and short period, J1906+0746 is a young pulsar. Its characteristic age
$\tau_c $ is roughly\,112 kyr, the lowest of all known binary pulsars,
and in the 8th percentile of the general pulsar age distribution
\citep{mhth05}.  Young pulsars show rapid spin-down evolution. By
definition this early 10$^5$-year stage is much more
fleeting than the 10$^9$-year detectable life span of {\it
  recycled} pulsars. In binary
systems the recycled pulsars are thus common, the
young pulsars rare. The recycled pulsars are generally the older of
the two binary components -- 
but the young pulsars formed more recently than their
compact-object companion.  Therefore these latter provide a
perspective on binary evolution that 
is different from the typical recycling scenario.

Pulsar J1906+0746 was intensively monitored with
the Arecibo, Green Bank, Nan\c{c}ay, Jodrell Bank and Westerbork radio
telescopes, up to 2009, after which the ever-decreasing pulse flux
density (Fig.~\ref{fig:profile_changes}) generally prevented
significant further detections. We here present the high-precision
follow-up timing from these telescopes over that initial five year
period 2005-2009.

Through this analysis we are able to significantly improve the system
characterization presented in the \citet{lsf+06} discovery paper: 
{
we measured two more post-Keplerian orbital parameters 
(for a total of three)} -- which,
assuming general relativity, provide an over-constrained
determination of the pulsar and companion masses.

In \S \ref{sec:1906:observations} we
report on the observing and basic data reduction methods.  Results on
a measurement of the distance to \J\ through its HI absorption are
presented in \S \ref{sec:1906:HI}.  In \S
\ref{sec:1906:profileevolution}-\ref{sec:1906:orbitalaberration} we
next detail our timing approach and methods, and the results from this
timing campaign. We discuss the implications of these measurements in
\S \ref{sec:1906:companion} and \ref{sec:1906:implications}.

\begin{center}
\begin{table*}[bt]
\centering
  \begin{tabular*}{0.92\textwidth}{ lcccccccr }
  \hline
  \hline
  {\bf Telescope}
                & {\bf Backend}
                  & {\bf Epoch}
                     & {\bf Cadence Interval}
                     & {\bf Central Frequency}
                        & {\bf Bandwidth}
                           & {\bf Channel BW}
                             & {\bf Coherent}
                               & {\bf N$_{\rm{\bf TOAs}}$} \\
               &      &           &            & {\bf (MHz)}    & {\bf (MHz)}  & {\bf (MHz)}& {\bf Dedispersion}  &   \\
  \hline
  Arecibo      & WAPP$^{1}$ & 2005-2009 & week-month & 1170,1370,1570 & 3$\times$100 & 0.195 & N  & 23250 \\
               & ASP$^{2}$  & 2005-2009 & week-month & 1420/1440      & 16$-$32      & 4     & Y  &   220 \\
  GBT          & GASP$^{2}$ & 2006-2009 & week-month & 1404           & 64           & 4     & Y  &  1110 \\
               & Spigot$^{3}$& -        & -     & 1440           & 600          & 0.781 & N  &  -    \\
  Nan\c{c}ay   & BON$^{4}$  & 2005-2009 & week  & 1398           & 64           & 4     & Y  &   650 \\
  Jodrell Bank & AFB$^{5}$  & 2005-2009 & 3-7\,days& 1402           & 64           & 1.0   & N  &  5010 \\
  WSRT         & PuMa$^{6}$ & 2006-2007 & month & 1380           & 80           & 0.156 & N  &    40 \\
               & PuMaII$^{7}$&2007-2009 & month & 1380           & 160          & 0.313 & Y  &    20 \\
	\hline
	\hline
	\end{tabular*}
	\caption{Details of the telescope and backend
          setup. N$_{\rm{TOAs}}$ marks the number of TOAs generated per
          backend. Spigot
          data was used for searching, not for 
          timing. 
$^{1}$\cite{dsh00};
$^{2}$\cite{dem07}, \cite{fer08};
$^{3}$\cite{kel+05};
$^{4}$\cite{des09};
$^{5}$\cite{hlk+04}; 
$^{6}$\cite{vkv02};
$^{7}$\cite{ksv+08}.
\label{table:obs}
}
\end{table*}
\end{center}

\section{Observations and initial data reduction}
\label{sec:1906:observations}

For the timing follow-up of this pulsar, we have obtained high
signal-to-noise data using the Arecibo Telescope and the \acf{GBT}, covering several full orbits 
with the latter.  High-cadence data from the Nan\c{c}ay, Jodrell Bank and
Westerbork telescopes provided further long-term timing coverage.

Data from the Arecibo telescope (USA) were taken using two backends
simultaneously, as detailed in Table~\ref{table:obs}.  Three
\acf{WAPP} filterbank machines autocorrelated the two polarization
channels. Offline, these were
converted to spectra, dedispersed incoherently, summed, and finally
folded at the local value of the pulsar period.  Further data were
taken using the \ac{ASP} coherent dedispersion machine, which folded
on-line using the best-known values for the dispersion measure and the
local pulse period. Through its
\textit{coherent} dedispersion 
capabilities, ASP is complementary to the WAPPs with their larger
bandwidths. In parallel to this on-line folding, \ac{ASP}
recorded the 4-MHz wide band around 1420\,MHz and stored the baseband
data on disk, for off-line investigation of HI absorption toward the
pulsar (\S \ref{sec:1906:HI}).

Data from the \ac{GBT} (USA) were collected with the Pulsar Spigot card and,
in parallel, the \ac{GASP} coherent dedispersion machine, which is a
GBT clone of \ac{ASP}.

At the Nan\c{c}ay
telescope (France), the \acf{BON} coherent dedispersion machine produced
dedispersed and folded profiles every 2 minutes.

Data taken with the Lovell Telescope at the Jodrell Bank Observatory
(UK) and its analogue filterbank (AFB) were dedispersed
and folded on-line at the nominal pulsar period.

At the Westerbork Synthesis Radio Telescope (WSRT) in The
Netherlands, data was initially recorded with the pulsar machine
PuMa. Starting in 2007, successor
machine PuMaII {added coherent dedispersion and} doubled the bandwidth (Table~\ref{table:obs}).

\subsection{Flux calibration and offline refolding}
\label{sec:1906:datareduction}

The WAPP, ASP and GASP data were flux-calibrated using the noise diode
signal that was injected into the receiver, for each polarization
individually, before each observation. 
When good calibration observations were not available, we normalized
the flux density in each profile by the root-mean-square (RMS) across
the profile for the coherently dedispersed profiles, while weighting
all channels equally for the WAPP filterbank data.  A continuum source
was used to further calibrate the ASP and GASP data for a significant portion
of the epochs, using ASPFitsReader \citep{fer08}. The \ac{WAPP} data were
calibrated using
\setcounter{footnote}{0}
\texttt{SIGPROC}\footnote{\url{http://sigproc.sourceforge.net/}} and
pre-recorded calibrator
data\footnote{\url{http://www.naic.edu/~phil/cals/cal.datR5}}. 

For all Arecibo and GBT data, time-averaged pulse profiles were
finally created by adding both polarizations, all frequency channels,
and five minute integrations.  These profiles were remade whenever
sufficient new Arecibo and GBT data were obtained to compute a new
ephemeris. As the coherently dedispersed ASP and GASP data were
recorded as a series of 30- or 60- second integrated pulse profiles,
these were subsequently realigned to create new 5-minute integrated
profiles.

The Jodrell Bank and Westerbork/PuMa profiles were produced using
their respective custom off-line software, while the Nan\c{c}ay and
the Westerbork/PuMaII profiles were reduced using the
\texttt{PSRCHIVE} software 
package\footnote{\url{http://psrchive.sourceforge.net/}} (\citealt{hvm04}),
These data were not flux-calibrated; they were normalized by the
\acs{RMS} of the noise.

\section{HI Absorption in the Pulsar Spectrum}
\label{sec:1906:HI}

\citet{lsf+06} combined the measured \ac{DM} of 218\,cm$^{-3}$\,pc with the NE2001 Galactic
electron-density model 
\citep{cl02} to estimate the distance to \J\ of \dtopulsarnoerr\,kpc. 
{Comparisons of NE2001 and VLBI distances suggest an error
of 20\% \citep[][although selection effects favoring 
  nearby pulsars may lead to much larger errors in several
  cases]{dtbr09}, producing an overall estimate of \dtopulsar\,kpc.}
Refining
that estimate would be beneficial for several lines of follow-up
analysis: it would improve our estimate of the kinematic contribution
to the observed orbital period decay $\dot{P}_b$, and thus strengthen
our \acl{GR} tests (\S \ref{pbdot_calc}). It would also influence the
likeliness of detecting a \ac{WD} companion (\S \ref{sec:1906:companion}).

Two more methods for distance determination exist, beyond the \ac{DM}
based method. These are based on parallax or kinematic measurements
\citep[for a in-depth comparison of these three methods,
  see][]{fw90}. 

Precise parallactic distances through \ac{VLBI} are now known {for} an
increasing number of pulsars 
\citep[e.g.][]{cbv+09}, and at a nominal distance of 5.4\,kpc,
\J\ is, in principle, nearby enough for such a
measurement. However, the measurement is best performed with in-beam
calibrators, and the process of identifying such sources was extremely
laborious during the period in which J1906+0746 was, at 0.55\,mJy
{mean flux density},
bright enough for observation with the Very Long Baseline Array
(VLBA). While the process of identifying in-beam calibrators has been
significantly streamlined at the VLBA with multi-phase center
correlation \citep{dbp+11}, \J\ had faded down to a mean
flux density of 33\,$\mu$Jy in 2012, making it unsuitable for VLBA
astrometry.

In the \emph{kinematic} method, the absorption of pulsar
emission by Doppler-shifted neutral hydrogen (HI) {along} the line of
sight, can, once combined with a model of the Galactic kinematics,
constrain the distance to the pulsar. {In the off-pulse phases}, the HI in its line of sight shines in emission;
but {during the on-pulse phases}, some of {the pulsar's} broadband emission
will be absorbed by
the HI clouds between it and Earth
(the ``pulsar off'' and ``pulsar on'' simulated spectra in
Fig.\ \ref{fig:hi}, left). 
As this absorption occurs at the
Doppler-shifted HI frequency, it can be associated with a velocity
relative to the observer. A Galactic rotation curve in the line of
sight can then model at which distances from Earth such a velocity is
expected. If the measured absorption velocities unambiguously map onto
the curve, a lower limit to the pulsar distance can be produced.  In
some cases, an upper limit on the pulsar distance can also be
derived: if certain features in the emission spectrum \emph{do
  not} have corresponding absorption dips, the pulsar can be assumed
to lie in front of the emitting region corresponding to the feature
velocity (see \citealt{fw90} and \citealt{wsx+08} for further discussion).

Deriving a kinematic distance to \J\ is especially interesting as its
\emph{DM-derived} distance is \dtopulsar\,kpc, near the tangent
point in this direction, at $6.4$\,kpc.  The absence of
absorption at the highest velocities, for example, would firmly put
\J\ closer to us than this tangent point.

Most kinematic HI
distances have been determined for bright, slow pulsars \citep[e.g.][]{fw90, jkww01}.
Of the 70 pulsars with kinematic distances
listed in \citet[][their Table 1, excluding distances derived from
  associations]{vwc+12}, \J\ is the 8th fastest spinning and, at
0.55\,mJy, the dimmest. Combined with the high DM, measurement of HI
absorption is challenging. 
We therefore used, for the first time in a pulsar HI
absorption measurement,
\emph{coherent dedispersion} to maximize the signal-to-noise ratio by
eliminating smearing between on- and off-pulse bins.

\subsection{Observations and analysis}
During four full tracks on \J\ with Arecibo, on 2006 June 14, July 11,
Oct 12 and November 12, we recorded a total of 7.6\,hr of baseband data with
\ac{ASP} (\S \ref{sec:1906:observations}). Using 
\texttt{DSPSR} \citep{sb11}, each of
the four observations {was} coherently dedispersed, folded onto 256
phase bins using the ephemeris resulting from the timing campaign
(Table \ref{table:1906timing}), and split into the maximum possible
number of channels of 1024 over
4\,MHz, for a velocity resolution of
0.83\,km/s. The long Fourier transforms needed for coherent dedispersion 
obviated further windowing functions. 

The two, main and interpulse, on-regions were
defined as the series of phase bins with \acl{SNR}\acused{SNR} SNR $>$ 3. The
off-pulse regions are the two stretches in between, minus 10-bin
margins. Within the off-pulse region, the spectra were
directly averaged over all bins to produce the off-pulse spectrum,
using \texttt{PSRCHIVE} \citep{sdo12}. For
the on-pulse spectrum, the spectra were averaged while weighting by
the square of the pulsar SNR in the concerning on-pulse bin 
{\citep[following][]{wsx+08}}. For each
observation, the channel frequencies were barycentered and converted
to velocities relative to the Local Standard of Rest (LSR).

\begin{figure}[tb]
   \centering
   \includegraphics[width=0.50\textwidth]{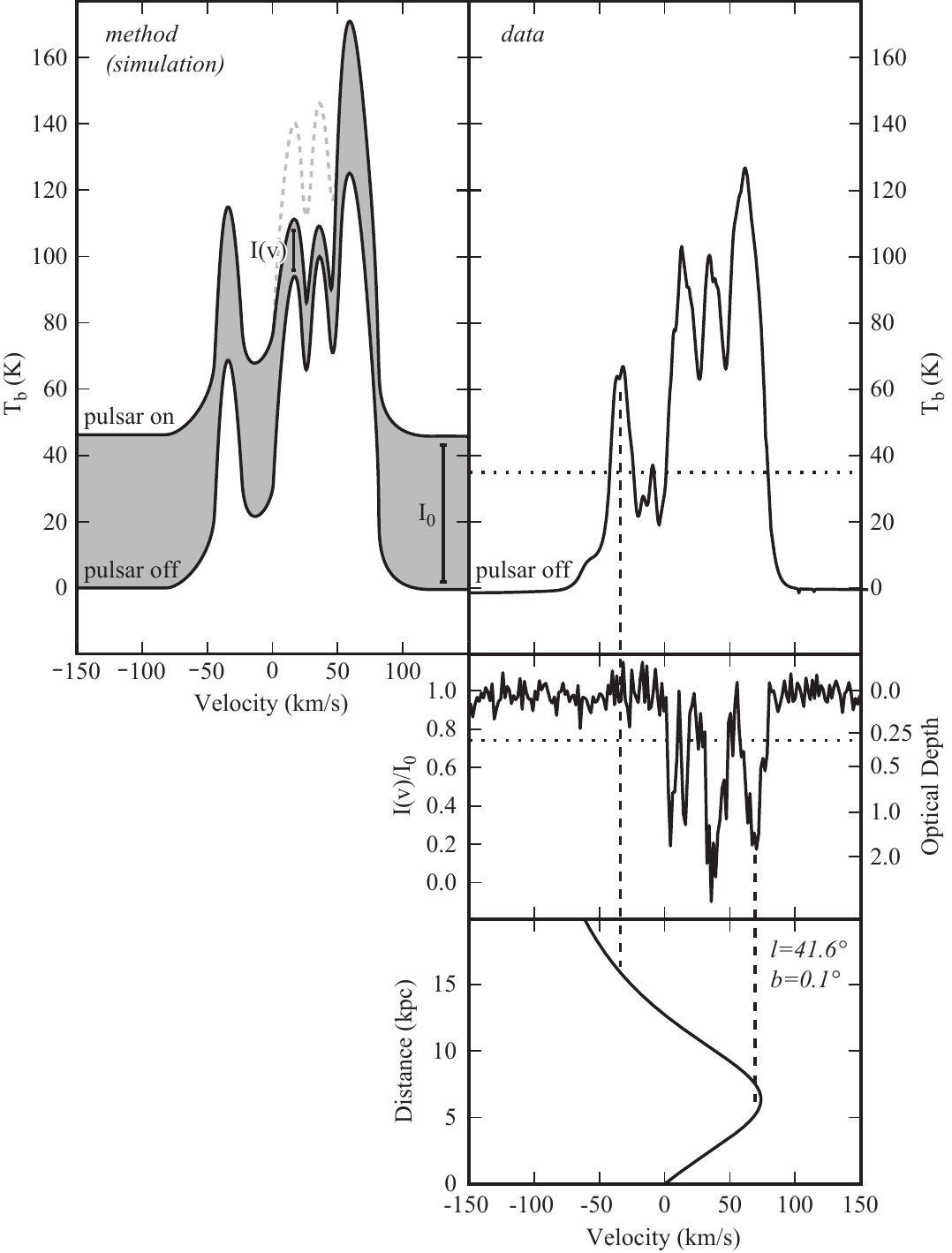}
   \caption{ Simulated (left) and measured (right) HI absorption data.
     Left: illustration of the subtraction and scaling methods used to
     define the absorption spectrum I(v)/I$_0$. A smoothed version of
     the \emph{measured} ``pulsar-off'' spectrum is shown as the
     bottom solid curve. The top, gray, dashed curve is based on the
     measured ``pulsar-on'' spectrum but the increase in intensity
     over the pulsar-off spectrum is \emph{amplified 100$\times$} for
     clarity, and shows the expected on-pulse spectrum if there were
     no absorption and $I(v)=I_0$. In the middle, solid, gray-filled curve,
     which then defines the absorption spectrum $I(v)/I_0$, absorption
     of the {two} features at 13 and 35\,km/s is simulated.
\\
     Right: The measured HI spectrum. In the top panel, the off-pulse
     HI emission spectrum is shown. The 35\,K criterion 
     {(\S \ref{sec:1906:hi:int})} is indicated 
     by the dashed horizontal line.  The middle panel contains the
     absorption spectrum $I(v)/I_0$.  The optical depth, derived from
     $e^{-\tau}$=$I(v)/I_0$, is noted with a dashed line at
     $\tau$=0.3. In the bottom panel, the Galactic rotation curve for
     this line of sight is plotted. The features in the absorption
     spectrum span the velocity range up to the tangent point. The
     derived lower and upper bounds to the distance are each marked
     with a vertical dashed line.}
   \label{fig:hi}
\end{figure}

The spectra from the four observations were summed, weighted by the
square of the pulsar \ac{SNR} in each observation, for both the on and
off-pulse. The intensity scales were calibrated by matching the peak
of the off-pulse spectrum to the peak brightness temperature $T_b$ as
measured in this direction by the VLA Galactic Plane
Survey \citep{std+06b}. 
Overall ``pulsar on''
and ``pulsar off'' spectra were thus produced.
Their difference $I(v)$, illustrated in Fig.\ \ref{fig:hi},
can be attributed to the pulsar {minus the absorption}.
By dividing by $I_0$, the broadband unabsorbed
strength of the  pulsar signal, the relative absorption
spectrum for \J\  was produced, as shown in the right-middle panel 
of Fig.\ \ref{fig:hi}. It shows several deep absorption features. 

\subsection{Interpretation}
\label{sec:1906:hi:int}
Any absorption features deeper than an optical depth $\tau$ of 0.3 {is
considered significant \citep[][]{fw90}}. Four of these appear, peaking at 4, 13, 35 and
63\,km/s.  To determine the distances to which these velocities
correspond, we constructed a Galactic rotation curve
\citep[the same as][]{vwc+12}
using a distance from the Galactic Center to the Sun of
$R_0$=8.5\,kpc and a flat rotation of $\Theta_0$=220\,km/s
\citep{fbs89}.

From this curve (Fig.\ \ref{fig:hi}, bottom panel), we
find that the highest-velocity emission component, at 63\,km/s, is
emitted near the tangent point.  This means the \emph{distance lower
  limit} is at or beyond that tangent point at 6.4\,kpc. 
By propagating the estimated spread in velocities of $\pm$7\,km/s
\citep{fw90}, we determine a distance uncertainty of 0.9\,kpc.

Any emission peak over $T_b$=35\,K is thought to be sufficiently
bright to allow for an absorption measurement \citep{wbr79}.  The
upper limit to the distance is thus provided by the farthest-out such
peak that is not accompanied by absorption.  The first peak to meet
these criteria is at $-$32\,km/s. This means the hard distance upper
limit derived using this kinematic distance is 15.9$\pm$0.9\,kpc, significantly
beyond the range allowed from the DM distance.

If the pulsar resides just at the tangent point, few features are
expected in the absorption spectrum, as there is only single chance
for absorption at a given velocity. If, however, \J\ is far beyond the
tangent point, there are 2 chances for absorption at each
velocity. The absorption spectrum of nearby pulsar B1904+06 shown in
\citet{cfkw88}, for example, is much more crowded than that of
\J. Overall this suggests the actual distance to \J\ is close to the
lower limit.

%%% LK bias and final result
\citet{vwc+12} present a sophisticated likelihood analysis of such measured
limits, to produce an overall distance estimate that takes selection
effects such as the {discovery} luminosity bias into account.
Using this method\footnote{\url{http://psrpop.phys.wvu.edu/LKbias}} 
we combine our upper and lower
limits with the discovery {mean flux density} of 0.55\,mJy \citep{lsf+06}, and obtain an
overall distance estimate of \HIdtopulsar\,kpc. That estimate
is larger than the previous, DM-derived distance of
\dtopulsar\,kpc.

\section{Profile evolution}
\label{sec:1906:profileevolution}
The profile of \J\ changes drastically over a timescale of
years.  The prominent interpulse in the 2004 discovery pulse profile
was absent from the 1998 Parkes archive profile
(Fig. \ref{fig:profile_changes}).
On the shortest time scales, our observations range from 
single-pulse data to 10, 30, 60 and 120-second integrations. From
visual inspection we have detected no mode 
changes on either of these timescales, or in the hours to weeks between
observations. 

We attribute the profile evolution to geodetic precession of the pulsar's
spin angular momentum vector about its total orbital angular momentum
vector. In \ac{GR}
 the precession rate \citep[e.g.][]{bo75} is:
\begin{equation}
\Omega_{\rm{geod}} = T_\odot^{2/3}\left(\frac{2\pi}{P_b}\right)^{5/3}\frac{1}{1-e^2} \ \frac{m_2\left(4m_1+3m_2\right)}{2\left(m_1+m_2\right)^{4/3}}
\label{eq:precession_period}
\end{equation}

\noindent where $T_\odot=G M_\odot / c^3=4.925490947$\,$\mu$$s$ is the
solar mass expressed in time units; $m_1$
and $m_2$ are the pulsar and companion masses, respectively, in solar
masses; $P_b$ is the orbital period, and $e$ is the eccentricity.
Using the timing solution and masses presented later, in Table
\ref{table:1906timing}, we find a predicted geodetic precession period of
$\sim$165\,years, which equals a rate of $2.2$ degrees per year.  Over
the 2005$-$2009 baseline described here, we thus expect a shift in
geometry of  roughly ten degrees.
The geometry has changed even more significantly -- by roughly
$20^{\circ}$ -- since the 1998 archival Parkes observation, 
qualitatively consistent with the significantly different pulse shape then.

The secular profile changes observed in \J\ offer an exciting
opportunity to study geodetic precession;
but the changing profile shape poses a problem in 
determining precise pulse times-of-arrival (TOAs).  
To ensure that the \emph{fiducial points} of all profiles are
consistent, and limit introduction of further timing noise,
we used a series of Gaussian standard profiles developed from the
well-modelled epochs of ASP, GASP and WAPP data, and next aligned these
Gaussian templates as described below.

\begin{figure}[bt] %  figure placement: here, top, bottom, or page
   \centering
   \includegraphics[width=\columnwidth]{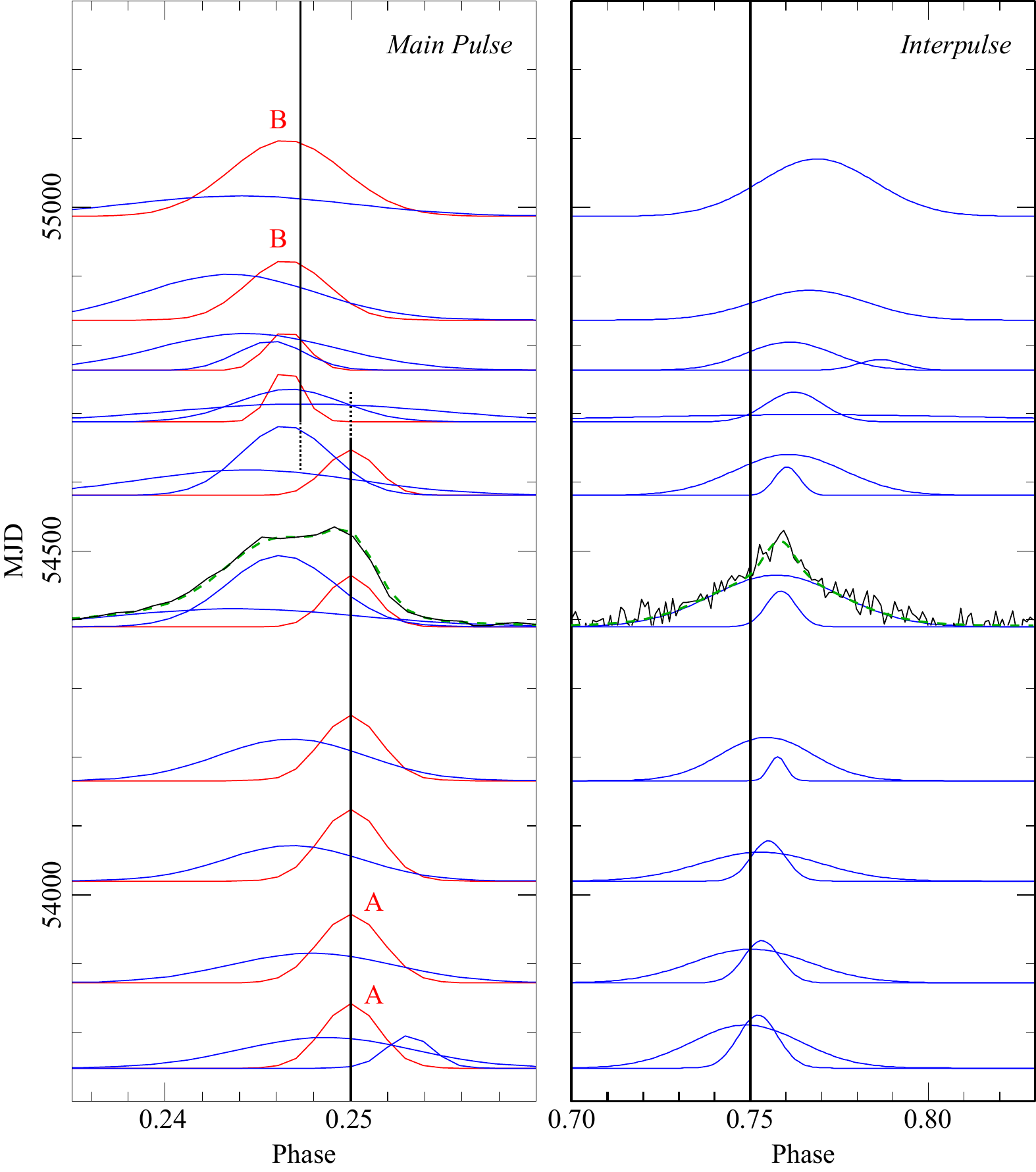}
   \caption[1906+0746: illustration of alignment of ASP and GASP
     Gaussian components]{
     Illustration of the alignment based on Gaussian components, for
     GASP and ASP profiles.
     Shown are the main pulse (left) and interpulse (right, flux scale
     increased 7 times).  The
     vertical line at phase 0.25 marks the location of the fiducial
     point.  The vertical line at phase 0.75 can be used to trace the
     changing location of the interpulse, over time.  The red lines
     represent the Gaussian component A or B that was used for the
     alignment. The blue lines are the other Gaussian components.
     For MJD 54390 the data (black line) and total profile model
     (dashed green line) are also shown.
}
   \label{fig:profevol_shift}
\end{figure}

\begin{figure}
  \includegraphics[width=\columnwidth]{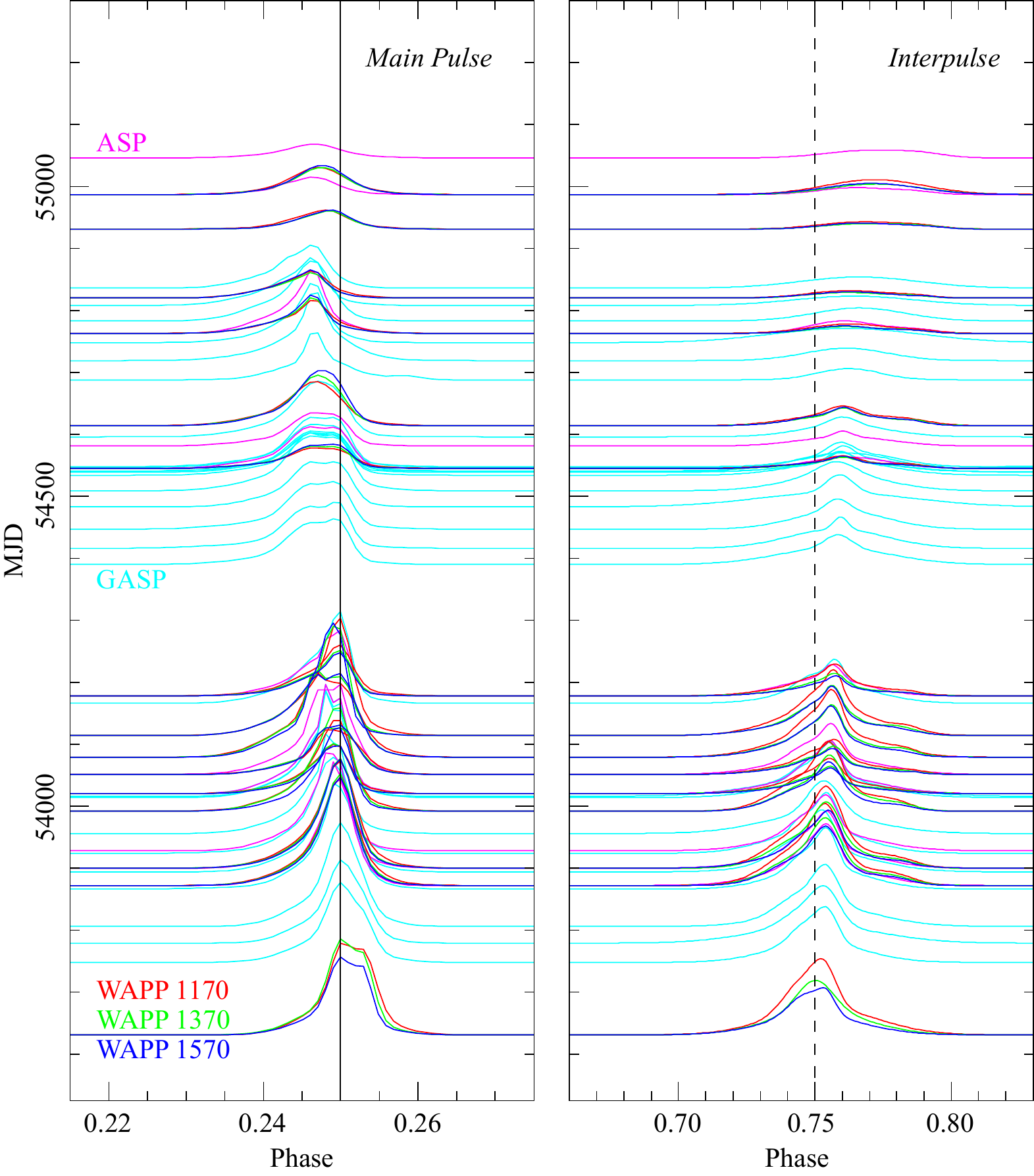}
  \caption[1906+0746: Full collection of Gaussian components]{The
    resulting fits
    to the pulse and interpulse of J1906+0746, based on data taken
    with the WAPPs (red: 1170\,MHz, green: 1370\,MHz, blue: 1570\,MHz)
    ASP (magenta) and GASP (cyan). The vertical axis spans from July
    2005 to August 2009. The interpulse flux is magnified by a
    factor of ten relative to the main pulse. The vertical line for
    the main pulse (left) illustrates the chosen alignment.
    A vertical line in the interpulse panel (right) illuminates the
    interpulse phase shift. 
}
   \label{fig:profevol}
\end{figure}

Using \texttt{BFIT} \citep{kwj+94} we fit sets of up to 3 (as
necessary) Gaussians to both the pulse and interpulse in the summed
profile of each epoch (illustrated for MJD 54390 in
Fig.\ \ref{fig:profevol_shift}).
This approach
was previously used to determine the geometry of the {PSR~B1913+16}
system \citep{kra98}. 
We next
identified a stable component, and used this component as our timing
fiducial point.  We found that the smoothest alignment was achieved by
keeping the phase of the initially tallest component (``Component A'')
constant.  Fig.\ \ref{fig:profevol_shift} shows this approach for a
subset of
the ASP and GASP summed profiles.  After \acf{MJD} 54700
we could no longer reliably identify Component A,
and so instead aligned the tallest component for {\it those} epochs
(Component B), and introduced a fiducial point phase shift based on
the transition profile in which components A and B were 
both identifiable (see Fig.\
\ref{fig:profevol_shift}). The full collection of Gaussian-modelled profiles is 
shown in Fig.\ \ref{fig:profevol}.  This method of aligning the
profiles produced fairly monotonic behaviour in the phase of
the interpulse, as can be seen by the gradual widening between
the interpulse peaks and the fixed, vertical dashed line
in Fig.\ \ref{fig:profevol}. After this alignments, these profiles were used as
the standards for high-precision timing.

\subsection{Profile flux variations}
Throughout the period covered by this study, the profile evolution of
\J\ was accompanied by a steady decrease in the pulsar mean flux
density (Fig.\ \ref{fig:profile_changes}). 
Independently processed, well calibrated,
ASP and GASP data produced mean flux density estimates that are consistent,
falling from 0.8\,mJy in 2006 to 0.2\,mJy in 2009 \citep[Fig 4.6 in][]{kasi11}.

\section{Timing}
\label{sec:1906:timing}

\subsection{Times of arrival}
\label{sec:timing:toas}

Times of arrival (\acsp{TOA}) were created for the profiles resulting
from the iterative data reduction process {described in \S
\ref{sec:1906:datareduction}}. For the WAPP Arecibo data, we created
separate profiles for each epoch and each frequency. 
The Gaussian-derived templates arising from ASP,
GASP and the three WAPPs are shown in Fig.\ \ref{fig:profevol}.  As
noted, timing standard profiles are commonly created with their main
peak as the fiducial point, at zero phase; here, in contrast, we
preserved the alignment of the Gaussian-derived templates, as is clear
from Fig.\ \ref{fig:profevol}.  For Arecibo, GBT and Nan\c{c}ay
timing, we used the evolving Gaussian templates described above (\S
\ref{sec:1906:profileevolution}). 
At Arecibo, there were unexplained offsets between the 3 WAPP machines
on MJD 53810, so that day is left out of the analysis.
For the higher-cadence Nan\c{c}ay
data, interpolation of the produced profiles provided a smooth
evolution of the components.
For Jodrell and Westerbork timing,
static standard profiles were created from bright, aligned subsets of
their respective observation campaigns.

\subsection{Timing solution for \J}
\begin{figure}[tb] %  figure placement: here, top, bottom, or page
   \centering
   \includegraphics[width=0.5\textwidth]{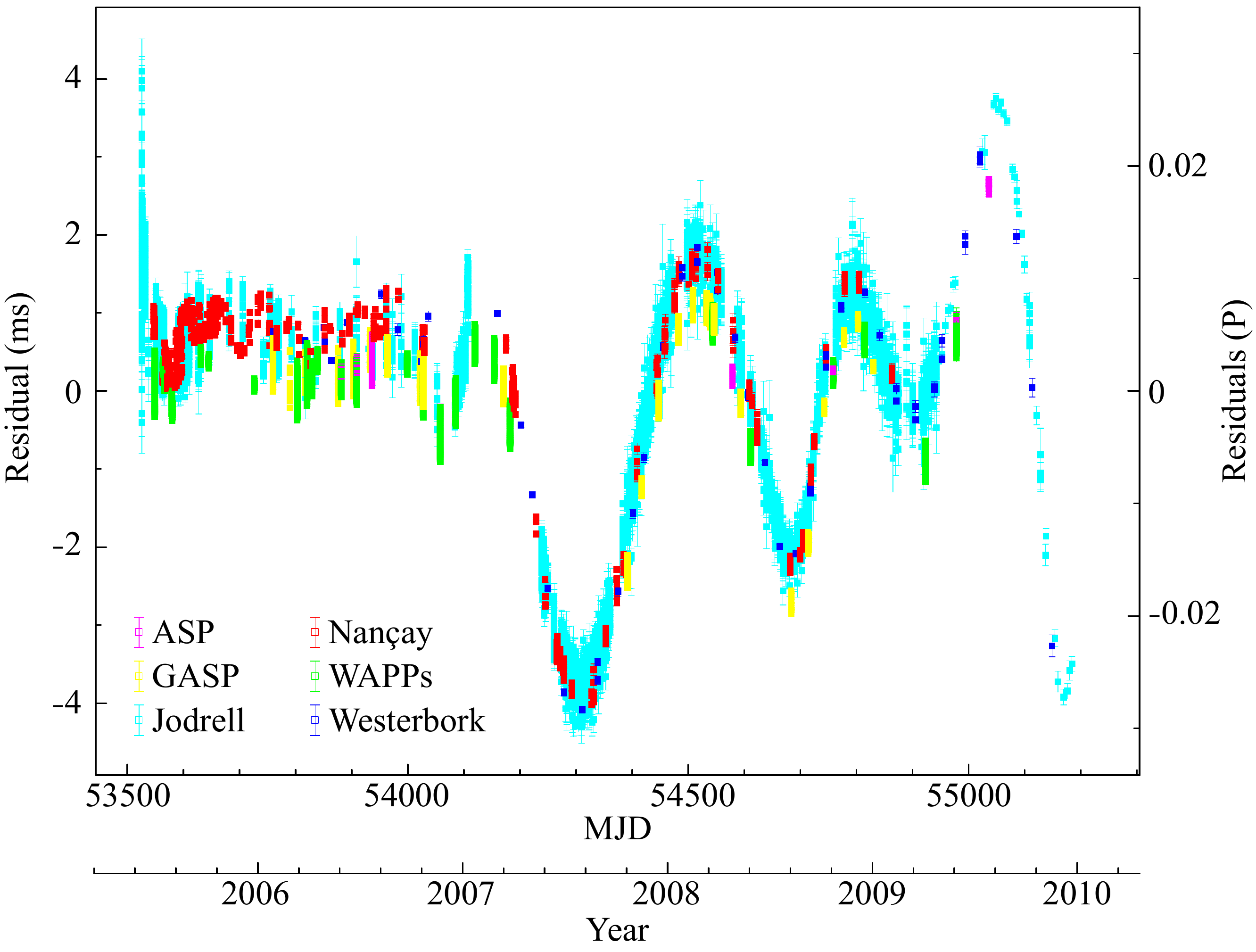}
   \caption[J1906+0746 All Telescope timing residuals]{Residuals plotted versus
     MJD and Year for the entire data set. Shown are the residuals
     after using the timing
     solution that included a 10th order polynomial in time. Also
     fitted were offsets between the various observatories, between
     pre- and post-54100 Jodrell data, and between PuMa and PuMaII. Data from
     ASP are magenta, GASP is yellow, Jodrell is cyan, Nan\c{c}ay is
     red, the combined WAPP data are shown in green, and Westerbork is blue.}
   \label{fig:1906:residuals_all}
\end{figure}

The complete set of 28,000 TOAs was fit for the parameters describing 
the state and evolution of both the individual pulsar and the binary system. This
initial fitting was performed using the \texttt{TEMPO2} \citep{hem06} package.
Using the data from  Arecibo, GBT, Jodrell, Nan\c{c}ay and Westerbork,
we produced a phase-connected solution over the entire period,
effectively accounting for every one of the 10$^9$ rotations over the 2005$-$2009 time span.
\J\ has a large amount of timing noise, however, that is difficult to
decouple from the orbital parameters. 
We first attempted to subtract this noise 
using \texttt{FITWAVES} (\citealt{hlk+04}), but better noise
removal was achieved by 
modeling the pulsar rotation frequency as a 10th order polynomial in
time, the highest degree of complexity 
currently implemented in  
\texttt{TEMPO2}. The {residuals of that timing solution are
  shown} in Fig. \ref{fig:1906:residuals_all}.

Even this last solution, however, shows large variations (of 420\,$\mu$s
rms) due to
remaining timing noise or unmodeled profile variations.
To
determine the system parameters with the highest precision, we thus
chose to
include only the timing data that used the evolving timing profiles,
{from Arecibo, GBT and Nan\c{c}ay (\S \ref{sec:timing:toas})}.  (The data from Jodrell
  and Westerbork 
were produced with static profiles and were well suited for inclusion in
the long-term high-cadence
timing solution but did not help the orbital timing solution.)
{In an attempt to limit the effects of timing noise and home in on the
best set of orbital parameters, we fit for an
arbitrary offset around 
{every set of TOAs derived  from a single observation}
 as well as individual-day DM
values on days with multi-WAPP Arecibo observations (see \S
\ref{sec:dmsecular}).  The
\texttt{TEMPO}\footnote{\url{http://tempo.sourceforge.net/}} code
allows up to a 20th-order polynomial in frequency, and we tested
fitting each of these orders while simultaneously fitting the offsets,
DM values, position and spin and binary parameters.  The fit that
allowed for a 4th-order frequency polynomial produced the lowest
reduced-$\chi^2$ and was adopted as our preferred timing solution.}

\begin{figure}[tbp] %  figure placement: here, top, bottom, or page
   \centering
   \includegraphics[width=0.5\textwidth]{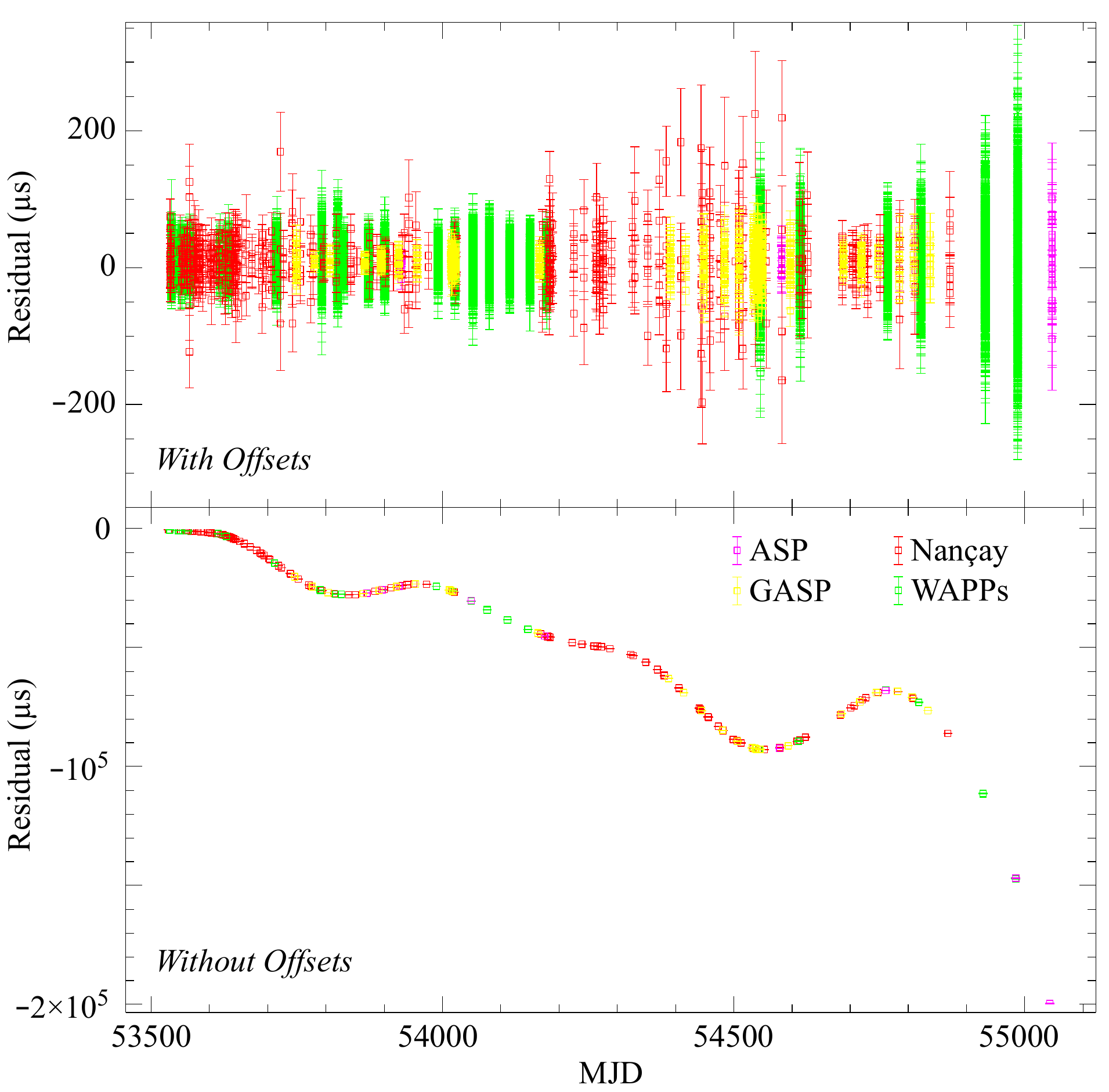}
   \caption[J1906+0746 Timing residuals]{Residuals are plotted versus
     MJD for the best fit. The top panel shows the residuals for the
% Nancay:118 (overlap with WAPP:25/each, ASP:9 GASP:28)
     solution that includes jumps between the 118 individual epochs. Those
     jumps remove the long-term timing noise trend that is clearly
     visible in the bottom panel. There we show the same
     phase-connected timing solution but without any jumps between
     epochs.  Green residuals represent WAPP Arecibo data, 
     magenta represents ASP Arecibo,
     yellow is GASP Green Bank, and
     red is Nan$\c{c}$ay data. \\}
   \label{fig:1906:residuals}
\end{figure}

The final, best timing solution is presented in Table
\ref{table:1906timing}, and the resulting residuals (17\,$\mu$s rms)
are shown in the
top panel of Fig.\ \ref{fig:1906:residuals}. 
We re-weighted our data so that the reduced
$\chi^2$ of the fit was equal to 1 for each data set, and overall.
Following common use, all reported uncertainties are twice the values
produced by \texttt{TEMPO} after this re-weighting.

\begin{center}
\begin{table*}
\centering
\begin{tabular*}{0.75\textwidth}{ p{78mm}  c c}
\hline
\hline
{\bf Measured Parameter} & {\bf DD Value} & {\bf DDGR Value}  \\
\hline
  Right ascension, $\alpha$ (J2000.0) \dotfill & \rightascension & \grrightascension\\
  Declination, $\delta$ (J2000.0)     \dotfill & \declination & \grdeclination \\
  Spin Period, $P$ (s)               \dotfill & \period & \grperiod \\
  Pulse Frequency, $\nu$ (s$^{-1}$)    \dotfill & \freq   & \grfreq   \\
  First derivative of Pulse Frequency,  $\dot{\nu}$  (s$^{-2}$)~ ($\times 10^{-13}$) \dotfill & \fo    & \grfo \\
  Second derivative of Pulse Frequency, $\ddot{\nu}$ (s$^{-3}$)~ ($\times 10^{-23}$) \dotfill & \foo   & \grfoo \\
  Third  derivative of Pulse Frequency               (s$^{-4}$)~ ($\times 10^{-30}$) \dotfill & \fooo  & \grfooo \\
  Fourth derivative of Pulse Frequency               (s$^{-5}$)~ ($\times 10^{-37}$) \dotfill & \foooo & \grfoooo \\
  Epoch (MJD)                             \dotfill & \epoch & \grepoch \\
  Dispersion Measure, $DM$ (cm$^{-3}$\,pc) \dotfill & \dm & \grdm \\
  Ephemeris                               \dotfill & \multicolumn{2}{c}{DE405} \\
  Clock                                   \dotfill & \multicolumn{2}{c}{TT(BIPM)}\\
  Orbital Period, $P_b$ (days)         \dotfill & \orbitalperiod & \grorbitalperiod \\
  Projected Semimajor Axis, $x$ (lt s) \dotfill & \semimajoraxis & \grsemimajoraxis \\
  Orbital Eccentricity, $e$            \dotfill & \eccentricity & \greccentricity \\
  Epoch of Periastron, $T_0$ (MJD)     \dotfill & \epochperiastron & \grepochperiastron\\
  Longitude of Periastron, $\omega$ (degrees)                      \dotfill & \periastron & \grperiastron \\		
  Rate of Periastron Advance, $\dot{\omega}$ (degrees/yr)          \dotfill & \omdot &\\
  Time Dilation and Gravitational Redshift Parameter, $\gamma$     \dotfill & \einstein & \\
  Orbital Period Derivative, $\dot{P}_b$ ($\times 10^{-12}$)        \dotfill & \pbdot & \\
  Excess Orbital Period Derivative, $\dot{P}_b$ ($\times 10^{-12}$) \dotfill & & \grxpbdot\\
  Total Mass, $M_{\rm{total}}$ ($M_\odot$)   \dotfill&  & \grmtotal \\
  Companion Mass, $m_{2}$ ($M_\odot$)      \dotfill&  & \grmcompanion\\
\hline	
\hline
{\bf Derived Parameter} & {\bf DD Value} & {\bf DDGR Value}\\
\hline
  Pulsar Mass, $m_{1}$ ($M_\odot$)      \dotfill& \mpulsar & \grmpulsar \\
  Rate of Periastron Advance, $\dot{\omega}$ (degrees/yr)      \dotfill &  & \gromdot \\
  Time Dilation and Gravitational Redshift Parameter, $\gamma$ \dotfill && \greinstein \\
  Orbital Period Derivative, $\dot{P}_b$ ($\times 10^{-12}$)     \dotfill & & \grpbdot \\
  Inclination angle, $i$ (degrees)    \dotfill & & \grinclination \\
  Galactic Latitude, $l$ (degrees)   \dotfill & 41.5982& \\
  Galactic Longitude, $b$ (degrees)  \dotfill & 0.1470&\\
  Mass Function, $f_{\rm{mass}}$         \dotfill & \massfunction & \grmassfunction \\
  Characteristic Age $\tau_c = P/2\dot{P}$ (kyr)   \dotfill &   \characteristicage & \grcharacteristicage \\
  Surface Magnetic Field, $B_{S} = 3.2\times 10^{19} (P\dot{P})^{1/2}$ ($10^{12}$ G)\dotfill& \surfB & \grsurfB \\
  DM-derived Distance to Pulsar, $d_{\rm{DM}}$ (kpc) \dotfill & \dtopulsar & \grdtopulsar \\
 HI-derived Distance to Pulsar, $d_{\rm{HI}}$ (kpc) \dotfill &  \multicolumn{2}{c}{\HIdtopulsar}\\
\hline
\hline
\end{tabular*}
\caption[J1906+0746 timing parameters]{Timing parameters for
  J1906+0746. The columns contain the
  parameters measured and derived using, on the left, the DD
  (\citealt{dd86}) model-independent timing model; on the right,
  the DDGR (\citealt{tay87}) timing model, which assumes
  general relativity to be the correct theory of gravity. 
  The
  DD model measures $\dot{\omega}$, $\gamma$, and $\dot{P}_b$,
  which can each be used to put constraints on the masses of
  the pulsar and companion.  The DDGR model measures the total
  mass $M_{\rm{total}}$ and the companion mass $m_2$ directly, and the
  post-Keplerian parameters can be derived from the values of
  the masses. The errors on these DDGR parameters
  $i$, $\dot{\omega}$, $\gamma$ and  $\dot{P}_b$ were
  derived through a Markov chain Monte Carlo 
  analysis based on the errors of  $M_{\rm{total}}$ and  $m_{2}$.
  Other errors reported here are \texttt{TEMPO} 2$\sigma$
  values.
  The DM-derived distance to the pulsar was
  estimated using the NE2001 model (\citealt{cl02}).%
% LEFT:  1906+0746.par.f4.3
% RIGHT: 1906+0746.par.f4.3.ddgr.2
%
}
\label{table:1906timing}
\end{table*}
\end{center}
\subsection{Measurement of post-Keplerian parameters}

Our timing solution presented in Table \ref{table:1906timing} includes
several post-Keplerian parameters. We describe these in some more
detail below.

\subsubsection{Measuring $\gamma$, the gravitational redshift}

% Now within three sigma
% $\gamma$ = \einstein\ ; $\gamma$ = 0.000493(8) in \citet{kasi07}

We tested how our measurement of $\gamma$ depends on the
current stage of the periastron precession cycle, which affects
the viewing geometry.
We simulated TOAs
over various points spanning one whole cycle, as was previously
shown for PSR B1913+16 and B1534+12, in \citet[][Fig.\ 5, top line]{dt92}. 
We find that our set of TOAs, collected while $\omega$ moved from $\sim 68^{\circ}$
to $78^{\circ}$, corresponds to relatively low theoretical fractional
uncertainties, and we are moving towards even better measurability
as $\omega$ increases \citep{kasi11}.

\subsubsection{Measuring $\dot{P}_b$, the orbital period derivative}
\label{pbdot_calc}
The observed value of $\dot{P}_b$ needs to be corrected for two
effects, before it
 can be compared to the value predicted by \acf{GR}, and used to
 constrain the system masses.
These are, first, the different Galactic acceleration felt by the
pulsar and by Earth; and second, the Shklovskii effect \citep{shk70},
which incorporates the Doppler effect caused by the pulsar proper
motion into the measured $\dot{P}_b$ value. 
We have no measurement of proper motion for this pulsar, but we
\textit{can} estimate
the Galactic contribution and calculate the limit on the Shklovskii
contribution and the system proper motion. If that limit is in
line with the measured proper motions of similar systems,  $\dot{P}_b$
can be used to constrain the binary system. 
Following \citet{dt91} and  \citet{nt95} we write the
observed orbital period decay $\dot{P}_b^{\rm{obs}}$ as:

\begin{equation}
\label{eqn:pbdot_obs}
\dot{P}_b^{\rm{obs}} = \dot{P}_b^{\rm{int}} + \dot{P}_b^{\rm{Gal}} + \dot{P}_b^{\rm{Shk}}
\end{equation}

\noindent where the Galactic contribution $\dot{P}^{\rm{Gal}}$ and
Skhlovskii term $\dot{P}_b^{\rm{Shk}}$ add to the intrinsic decay
$\dot{P}_b^{\rm{int}}$. 
If we 
assume that $\dot{P}_b^{\rm{int}}$ is equal to $\dot{P}_b^{\rm{GR}}$, the value determined by GR, we can rewrite Eq.\ \ref{eqn:pbdot_obs} to isolate
the Shklovskii contribution, scaled to the binary
orbit:
 
 \begin{equation}
\label{eq:pshk}
 \left(\frac{\dot{P}_b}{P_b}\right)^{\rm{Shk}} =  \left(\frac{\dot{P}_b}{P_b}\right)^{\rm{obs}} -  \left(\frac{\dot{P}_b}{P_b}\right)^{\rm{GR}} - \left(\frac{\dot{P}_b}{P_b}\right)^{\rm{Gal}}
 \end{equation}

On the right-hand side of Eq.\ \ref{eq:pshk}, each of the terms can be
estimated or 
calculated. First, our best fit value of the orbital decay
$\dot{P}_b^{\rm{obs}}$ is \pbdot $\times$10$^{-12}$
(Table~\ref{table:1906timing}).

Second, 
the value predicted by GR and computed
from fitting the DDGR model \citep{dd86,tw89} to
our data \mbox{$\dot{P}_b^{\rm{GR}}$ is \grpbdot$\times$10$^{-12}$}
(Table~\ref{table:1906timing}).

The third term, the Galactic contribution, can be written as
$(\dot{P}_b/{P_b})^{\rm{Gal}} = {\vec{a}\cdot\hat{n}}/{c}$, where $\vec{a}$ is the differential acceleration in the
field of the galaxy and $\hat{n}$ is the unit vector along our line of
sight to the pulsar. The components parallel and perpendicular to the
Galactic plane that make up this term
can be calculated \citep[][ Eqs.\ 3$-$5]{nt95}.
Using (again) $\Theta_0 = 220$\,km/s and $R_0 =
8.5$\,kpc, plus the pulsar coordinates and HI-absorption distance
from Table~\ref{table:1906timing}, we find 
$(\vec{a}\cdot \hat{n}/c)_{\parallel} = (6.3\pm2.5) \times
10^{-19}$\,s$^{-1}$ and
$(\vec{a}\cdot \hat{n}/c)_{\perp} = (4.5\pm1.1) \times 10^{-23}$\,s$^{-1}$.
These translate to a Galactic correction
$\dot{P}_b^{\rm{Gal}} =$ 0.009(4)$\times$10$^{-12}$.
Given the large distance uncertainty, this value does not significantly
change when using the more recent \citet{rmb+14} Galactic 
kinematics. 
Combined, the three terms limit the Shklovskii contribution to
be essentially zero,
${\dot{P}_b}^{\rm{Shk}} = {P_b}\mu^2{d_{\rm HI}}/{c} <
0.03 \times 10^{-12}$ (at 95\% confidence level),
 where $\mu$ and $d_{\rm HI} $ are the total proper motion and the
distance of the pulsar respectively.
The error ranges on ${P_b}$ and ${d_{\rm HI}}$ allow for a proper
motion of $< 9$\,mas/year or transverse velocity of  
$v = \mu\,d_{\rm HI} < 400$\,km/s (95\% CL).  
This easily encompasses the range of published system velocities for
other relativistic pulsars (e.g. \citealt{hllk05}).

Our results thus imply that the orbital period decay $\dot{P}_b$ for
J1906+0746 is consistent with the value predicted by general
relativity.

\subsection{Mass measurements}
Having obtained reasonable estimates of the advance of periastron
$\dot{\omega}$, the gravitational redshift/time dilation parameter
$\gamma$, and the orbital decay $\dot{P}_b$ for J1906+0746, we use
these three parameters to place constraints on the masses of the
pulsar ($m_1$) and companion ($m_2$).

\begin{figure}[tb] %  figure placement: here, top, bottom, or page
   \centering
   \includegraphics[width=0.5\textwidth]{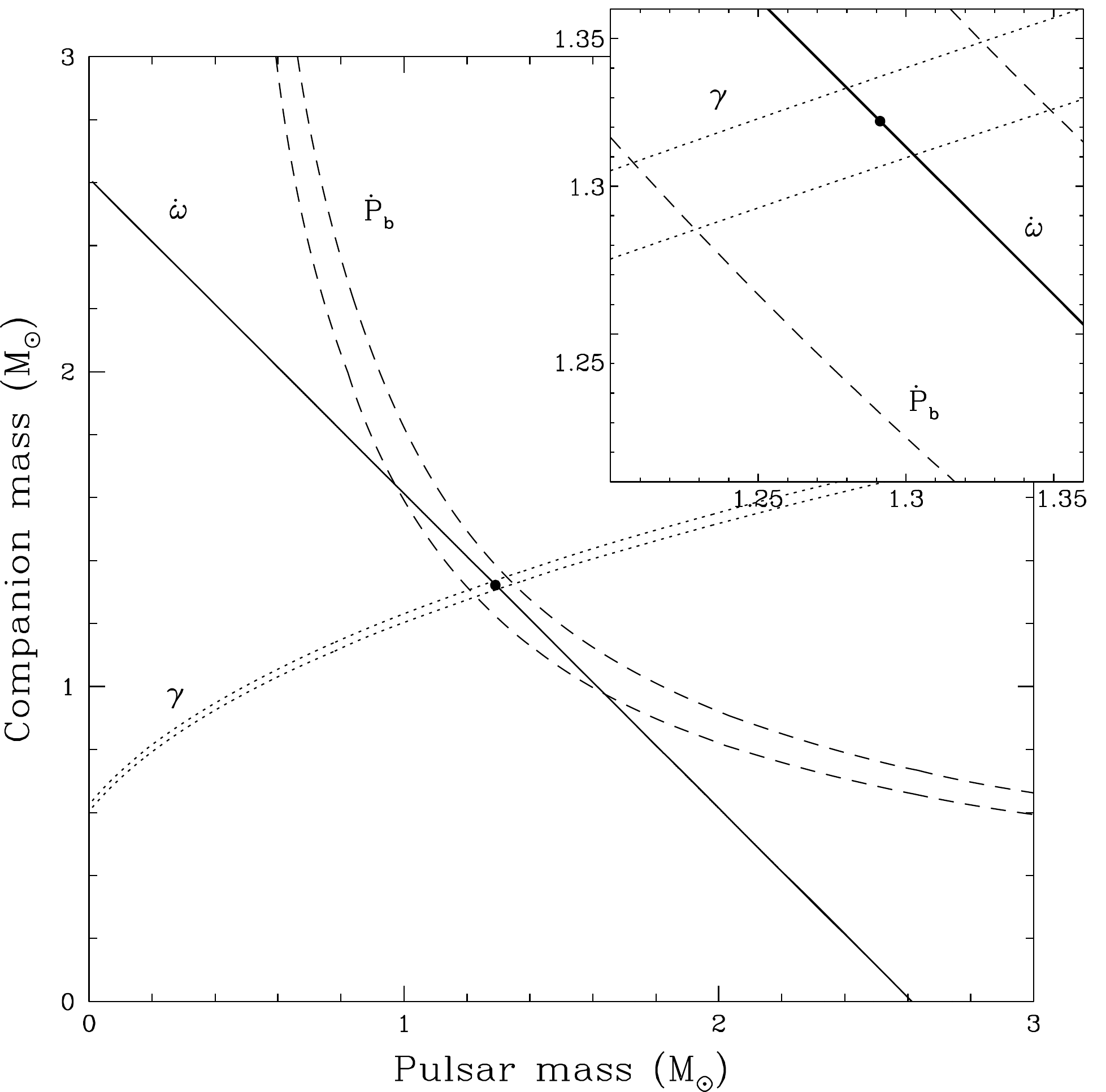}
   \caption[1906 mass-mass diagram]{Mass-mass diagram for the
     ephemeris presented in Table \ref{table:1906timing}. The lines
     represent the values of $m_1$ and 
     $m_2$ allowed by the three measured DD post-Keplerian parameters:
     $\dot{\omega}$ (the solid line), $\gamma$ (the dotted lines) and
     $\dot{P}_b$ (the dashed lines). The dot indicates the best-fit
     value for $m_1$ and $m_2$.
   }
   \label{fig:mass_mass}
\end{figure}

If we use the dependence of the post-Keplerian parameters on the
masses, as defined in general relativity \citep[see, e.g.\, ][]{tw89},
each parameter constrains the allowed ($m_1$,$m_2$) pairs.  
The intersection of the allowed regions in \mbox{$m_1$ - $m_2$} parameter
space represents the most likely 
values of the pulsar and companion masses.  The mass-mass diagram for
our timing solution of J1906+0746 is shown in Fig.\
\ref{fig:mass_mass}.

\noindent The three measured
post-Keplerian parameters provide consistent values of the pulsar and
companion masses. 
Using the DDGR (\citealt{dd86}, \citealt{tw89})
binary model, we finally conclude that the pulsar mass $m_1 =$ \grmpulsar\,$M_\odot$ and
the companion mass $m_2 =$ \grmcompanion\,$M_\odot$. 
That pair of masses  is marked in Fig.\
\ref{fig:mass_mass}, and indeed  falls within the overlap of the
constraints from the post-Keplerian parameters.

These masses differ from the initial estimates reported in \citet{kasi07}, at
$m_p=1.248(18)M_\odot$ and $m_c=1.365(18) M_\odot$ for the pulsar and
companion respectively. This change
 can be explained by the longer data span used here, and by
the improved method of eliminating the strong long-term timing noise.

~\\
\section{Dispersion measure variations}
We next investigate changes in dispersion measure,
either as long-term evolution, or as trends that could recur every orbit.
To estimate such variations within our data span,
we used the WAPP data, with its superior wide bandwidth and \mbox{512-channel}
spectral information. 

\subsection{Secular DM variation} \label{sec:dmsecular}
\begin{figure}[b] %  figure placement: here, top, bottom, or page
   \vspace{2mm}
   \centering
   \includegraphics[width=0.5\textwidth]{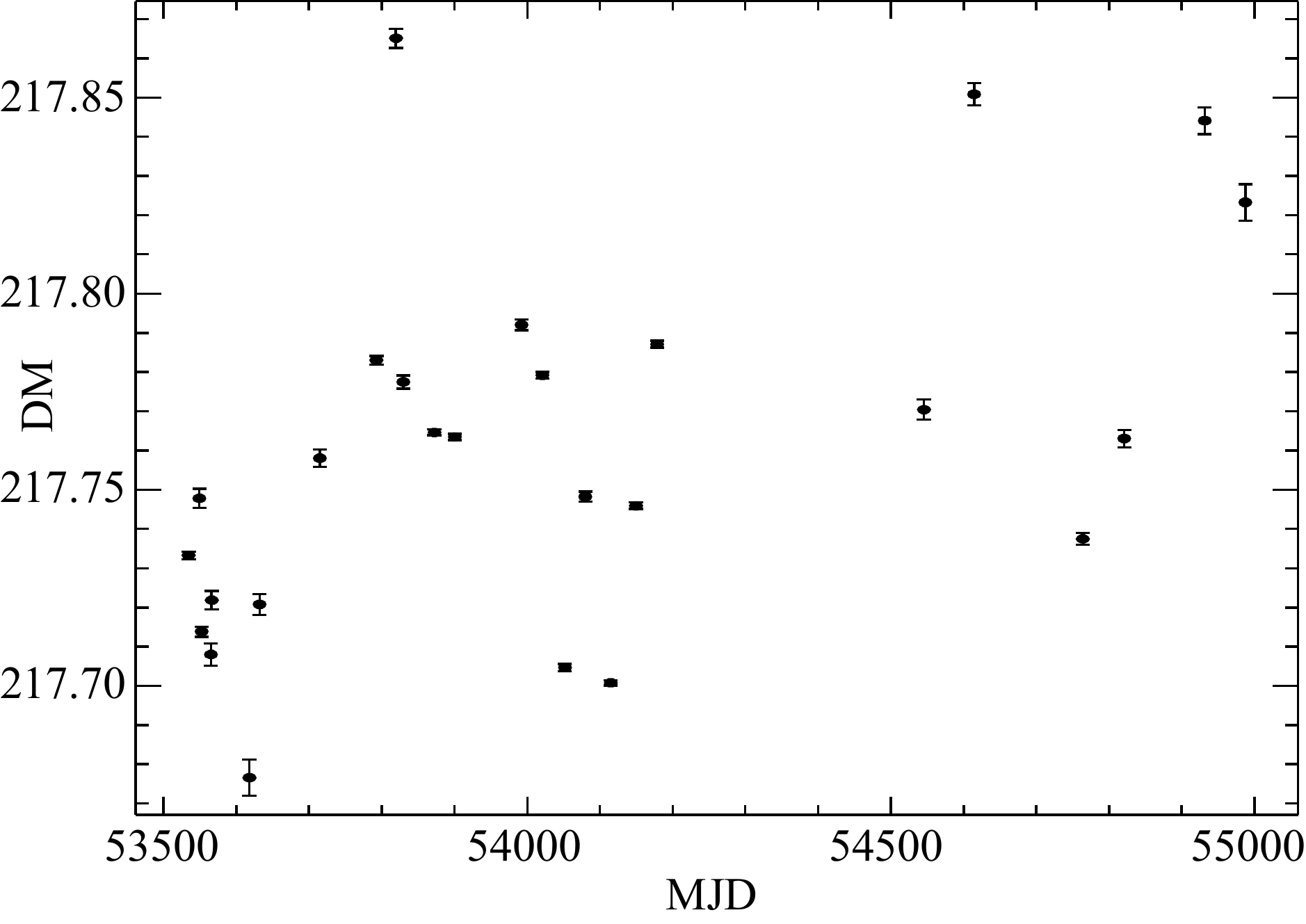}
   \caption[DM vs MJD for all epochs where WAPP data were available for
     J1906+0746.]{Dispersion measure variations versus time, for all
     WAPP-data epochs, fit simultaneously with the rest of the timing
     model.
   \label{fig:dm_vs_mjd}
}
\end{figure}

For each epoch of WAPP data with 3 WAPPS, we used \texttt{TEMPO} to fit for the DM
using only TOAs from that day.  We accomplished this by using the
``DMX'' model within \texttt{TEMPO}, fitting the individual-day DM values
simultaneously with the rest of the timing solution. 
This can bring to light intrinsic DM variations, but
it can also absorb other frequency dependent effects.  Note that the
ASP data at these epochs were stil fit with surrounding offsets,
because they did not in general agree with the WAPP DM values.  This
is presumably due to the fact that the ASP profiles were obtained with
coherent dedispersion but also folded in real-time with some
unavoidable ephemeris smearing.
Generally, the 
observed
variations are larger than the error bars.
There is,
  however, no long-term behaviour that could 
be included in the timing analysis (Fig.\ \ref{fig:dm_vs_mjd}). The short-term variations observed
in the DM values are most likely induced by other TOA changes that
systemically depend on frequency; these could be profile evolution
that varies with frequency, potentially amplified when scintillation
affects the relative contributions of different parts of the band.

\subsection{Orbital DM variation}
The dispersion measure may vary with orbital phase,
if the pulsar emission travels through varying plasma densities throughout the binary
orbit. Such variations have been seen in several pulsars with
non-degenerate companions (e.g., B1957+20, \citealt{fbb95}; and
J2051$-$0827, \citealt{svbk01}). To
distinguish between a neutron star and a less compact companion, we
investigated the behaviour of the DM of J1906+0746 over orbital phase.

\begin{figure}[tb] %  figure placement: here, top, bottom, or page
   \centering
   \includegraphics[width=0.5\textwidth]{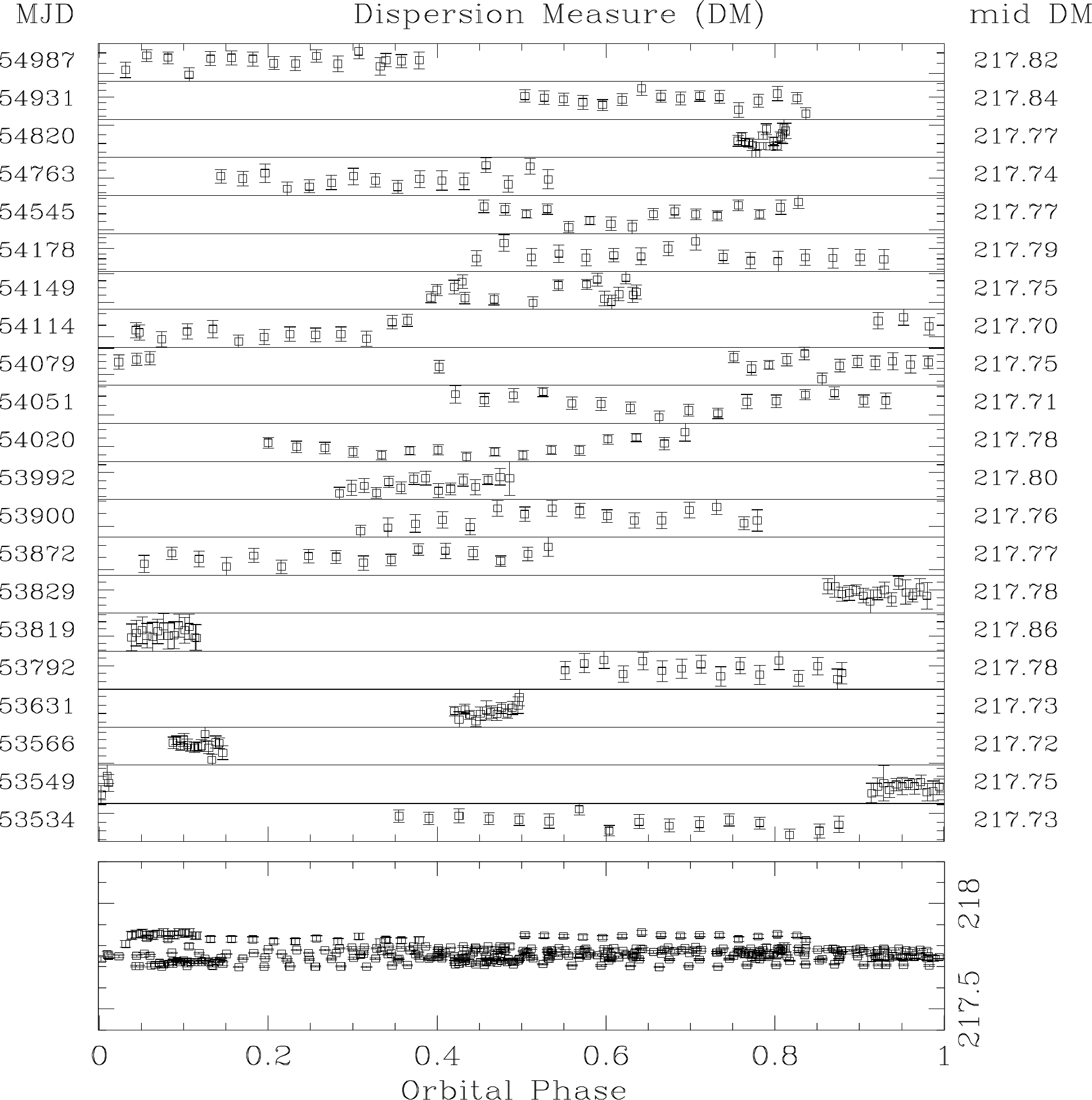}
   \caption[DM versus orbital phase for all epochs where WAPP data
     were available for J1906+0746.]{DM versus orbital phase for all
     epochs where good WAPP data were available for J1906+0746, folded
     on the \citet{kasi11} ephemeris.
     The bottom plot shows the DM over orbital phase for all of the epochs together. }
   \label{fig:dm_vs_orb_phase_all}
\end{figure}

For each WAPP-data epoch we divided TOAs over 16 bins across the
$\sim$2\,hr Arecibo observation. We calculated the best-fit DM for each of
the 16 bins individually, keeping other timing parameters fixed at the 
best-fit values  (Table~\ref{table:1906timing}).  We then
investigate the DM versus orbital phase for each epoch, as
plotted in Fig.\ \ref{fig:dm_vs_orb_phase_all}.  Although
variations at individual epochs are somewhat significant within the
stated, \texttt{TEMPO}-doubled error bars, there
are no compelling overall trends in DM over the course of an
orbit, and no evidence for a more extended companion.

\section{Orbital aberration}
\label{sec:1906:orbitalaberration}
We have so-far attributed the observed profile changes to
\textit{geodetic precession}, an effect seen only in strong
gravitational fields. It is, however, also possible that the special
relativistic effect of \textit{aberration} contributes to the observed
profile changes.  Aberration on an orbital timescale arises from the
relativistic velocities with which the pulsar and companion travel in
their orbit \citep{dt92,rl06}, and was measured in the double neutron
star B1534+12 \citep{sta04,2014ApJ...787...82F}.  As the velocity at periastron of
J1906+0746, $v_p \sim 0.001c$, is similar to that of B1534+12
\citep[Eq.\ 8.35;][]{lk05}, one might expect to detect aberration.
A {\it longitudinal} delay
will shift the pulse profile in phase, while keeping the shape of the
pulse intact; meanwhile, a {\it latitudinal} delay shifts the observed
emission angle with respect to the pulsar spin axis
\citep{rl06}. That latter change in our line of sight could produce measurable
profile changes over the course of a binary orbit. 

If detected, we can combine the profile changes due to orbital
aberration with the secular changes due to geodetic precession to put
limits on the geometry of the system (\citealt{sta04}).  Fitting the
polarimetry data for J1906+0746 to the classical
\ac{RVM} resulted in
the following measured angles \citep{des09}:
% (assuming an inclination angle of $i =
%43^{\circ}$, from timing results presented in \citealt{kasi07}): 
the current angle between the spin and magnetic axes
$\alpha = 80^{\circ +4}_{-6}$; the geodetic precession phase
$\Phi^0_{SO} = 109^{\circ +51}_{-79}$; and the precession cone opening
angle (or misalignment angle) $\delta=110^{\circ +21}_{-55}$.  A
detection of orbital aberration could constrain further angles and
would allow for a measurement of the geodetic precession period
independent of the profile beam model.
%, which would provide a
%consistency check. 

\subsection{Observations} % 
The detection of profile changes on an \textit{orbital} timescale,
against a backdrop of steady profile evolution from geodetic
precession, requires several complete orbits of coverage collected
over a relatively short time span.
We thus collected \ac{GBT} \ac{GASP} data during two separate,
high-cadence campaigns - over four days between October 4 and 12,
2006, and over 14 days between March 9 and 23, 2008. These campaigns
(cf.\ \S \ref{sec:1906:observations}) are each much shorter than the
$\sim$165\,year geodetic precession period.
The fully steerable GBT allowed tracks of the pulsar for
at least one full orbit per day, in both campaigns.

\begin{figure*}[tb]
\includegraphics[width=\textwidth]{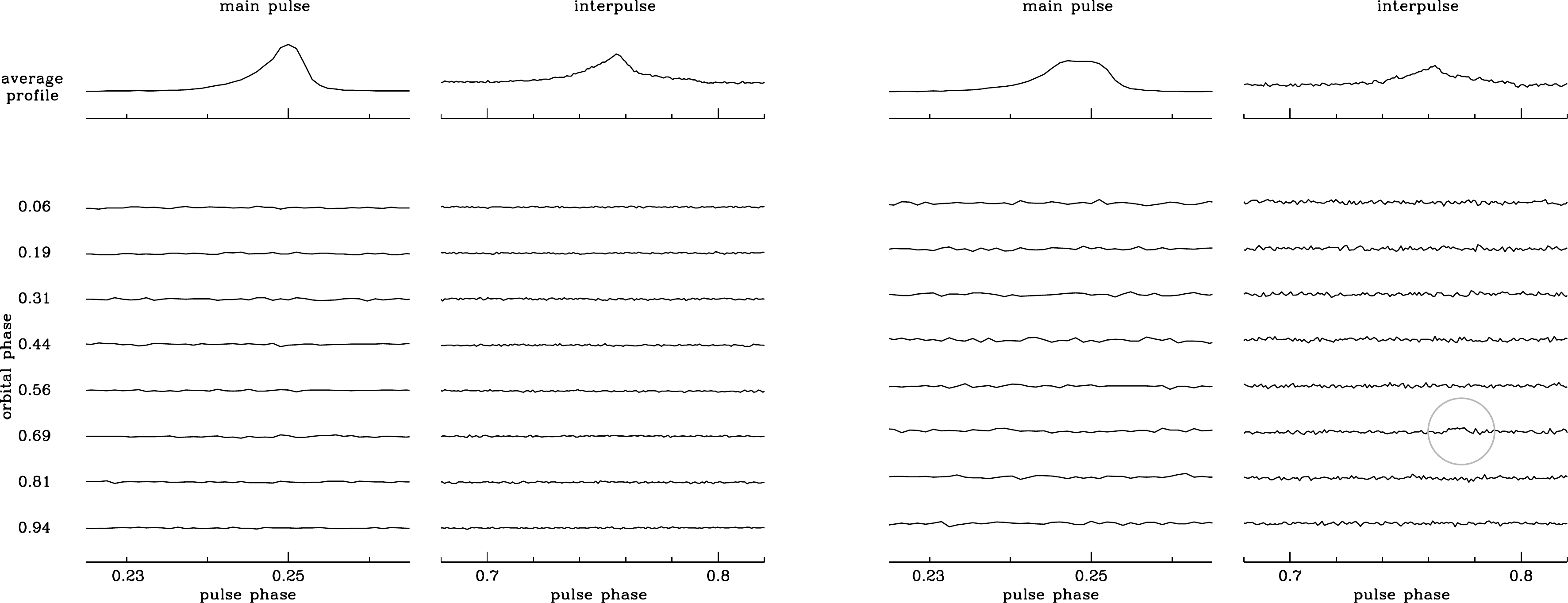}
  \caption[J1906+0746 - orbital aberration search over 8 orbital
    bins]{The pulse and interpulse average profile of 
    J1906+0746 (top row), and the difference between these total
    averages, and the average for 8
    orbital phase bins (bottom rows). Data were taken over four days in October 2006
    (left) and three days in March 2008 (right).
    The interpulse flux has been magnified 8-fold relative to the main
    pulse. The top row shows the average profiles per campaign.
    The subpanels below shows 
    the difference of the profile at orbital phase compared to the average.
 \label{fig:aberration1}}
%\end{figure}
\end{figure*}

\subsection{Method}
\label{sec:method}
To measure pulse profile changes versus orbit, separately for each
campaign, we used the following
approach, adapted from \citet{2008AIPC..983..474F}.
We produced time-averaged pulse profiles per five minute interval,
using the ephemeris derived by  \citet{kasi11}. From these,
we first created an average profile for the entire campaign.  
Each
five-minute profile
was next binned by true anomaly, over 8 orbital phases.
After scaling all profiles
to a uniform height, the bins were summed.
We then investigated the differences, shown in Fig.\ 
\ref{fig:aberration1},  between the 
total profile and each binned profile.

{For both the interpulse and the main pulse, we subtracted a
  linear baseline over their respective data window (the horizontal
  range in the subplots of Fig.\ \ref{fig:aberration1}). We then
  looked for series of 5 subsequent time bins that were offset from
  the mean by more than 1 standard deviation.}

\subsection{Results}
{Only one such instance was found, indicated with the gray
  circle in Fig.\ \ref{fig:aberration1}, but no related changes were
  detected at other orbital phases.
Thus no variations on orbital timescales were
detected with this method.} The difference profiles, shown in
  Fig.\ \ref{fig:aberration1}, do not 
show significant other changes over the course 
of an orbit. From this we conclude that 
J1906+0746 has a small aberration amplitude; and 
that the  long-term profile evolution can be used to constrain the
emission beam. % \citep[e.g.\ ][]{dkc+12}.

\section{Nature of the companion}
\label{sec:1906:companion}
From our timing campaign we find a companion mass
$m_c$=\grmcompanion\,$M_\odot$ and a pulsar mass
$m_p$=\grmpulsar\,$M_\odot$.  As evident from Table
\ref{table:binaries}, these masses well fit the observed
collection of \acp{DNS}, and the standard model for \ac{DNS}
evolution, in which the recycled companion is more massive than - or
at least comparable to - the young neutron star.
The mass of the companion is, however, also similar to that of the
massive WD in the relativistic binary with young pulsar B2303+46
(Table \ref{table:binaries}).  Thus, the masses alone cannot rule out
the companion is a \ac{WD}.  

For some binary pulsars it is possible to
observe the \ac{WD} companion optically; however, as discussed in
\citet{lsf+06}, this is not a viable option for J1906+0746.  As
detailed in the next section, we would expect a WD companion to have
an age of at least $\sim$1 Myr. Now, older and equally distant white
dwarfs have been optically detected \citep[B2303+46;][]{vk99}; but for
the low Galactic latitude of \J\ ($b$=$0.15^\circ$) the modeled
extinction A$_V$ out to $d_{\rm{HI}}$=7.4\,kpc can range from 4.1--8.4
\citep[model A--S in][]{al05}. That suggests a V magnitude of at least
29, too faint to reasonably detect.  Optical studies of the companion
can therefore not confirm or rule out its nature.

If \J\ is part of a double neutron star binary, it may be possible to
detect 
the companion as a second radio pulsar -- as seen in the double pulsar
system, J0737$-$3039A/B \citep{lbk+04}. We have therefore performed
searches for pulsed emission from the companion in
the WAPP and Spigot data. 
A \texttt{SIGPROC}
time series, dedispersed at the DM of the known pulsar,
was transformed into the companion rest frame for a range of
possible projected semimajor axes. We then searched for periodicities
in these data, and folded the transformed time series at the candidate periods.
No convincing pulsar signals were found. If the
companion is a pulsar, it is either
beamed away from Earth; or too dim.
The minimum flux density  $S_{\rm{min}}$ that we could have detected was:
\begin{equation}
S_{\rm{min}} = \beta\frac{\sigma_{\rm{min}}\left[T_{\rm{sys}}+T_{\rm{sky}}(l,b)\right]}{G\sqrt{n_p t_{\rm{obs}}\Delta\nu}}\sqrt{\frac{W_e}{P-W_e}}
\end{equation}
\noindent (\citealt{dic46,dtws85}) where $\sigma_{\rm{min}} = 8$ is the
threshold detection \acf{SNR}; $n_{p} = 2$ is the
number of summed polarizations, $\Delta \nu$ is the
bandwidth (600\,MHz with the Spigot; 3$\times$100\,MHz for the WAPPs), $\beta  \sim 1.2$ is the
quantization factor for 3-level quantization
\citep{lk05}, $G$ is the antenna gain (2.0\,K/Jy for
the L-band receiver at the
GBT\footnote{\url{http://www.gb.nrao.edu/~fghigo/gbtdoc/sens.html}} and
10\,K/Jy for the L-wide receiver at
Arecibo\footnote{\mbox{\url{http://naic.edu/~astro/RXstatus/Lwide/Lwide.shtml}}}),
$T_{\rm{sys}}$ is the system temperature (20\,K and 25\,K for the GBT
and Arecibo, respectively), $T_{\rm{sky}}(l,b)$ 
is the temperature of the sky at the location of the source
(\citealt{hssw82}), $W_e$ is the effective pulse width of the pulsar,
and $P$ is the pulse period.  For an integration time of
$t_{\rm{obs}}\sim$8 hours with Green Bank, the longest observation we
searched, and assuming an effective pulse width of 10\% 
our flux density limit $S_{\rm{min}} \simeq$19.7\,$\mu$Jy.
For our $\sim$2 hour observations with Arecibo we reach
$S_{\rm{min}}\simeq$19.3\,$\mu$Jy.  
At the HI-absorption distance of 7.4\,kpc,
any companion pulsar beamed toward us thus has a 
pseudoluminosity $S_{1400}d^2$ $<$
1.1\,mJy\,kpc$^2$.
Comparing this with the L-band pseudoluminosities 
of the \emph{recycled} pulsars in
known double neutron stars
% (6.5\,mJy kpc$^2$ for J0737$-$3039A, 
% 1.6 * 1.1^2 = 1.9
(1.9\,mJy kpc$^2$ for J0737$-$3039A, 
% 0.6 * 1.0^2
 0.6\,mJy kpc$^2$ for B1534+12,
% 0.9 * 7.13^2
and 45\,mJy kpc$^2$ for B1913+16 -- \citealt{bjd+06, dbt09,
  kxl+98,tc93}) we conclude that our search {would have
  detected 2 out of 3} of these at the HI distance of \J, and thus had
sufficient sensitivity to detect the average known recycled pulsar in
a \ac{DNS}, if its beam intersected Earth.

If the opening angle between the spin axis of this neutron star
companion and the angular momentum of the orbit is large enough, the
putative recycled pulsar will become visible within a geodetic
precession timescale (Eq.~\ref{eq:precession_period}). Continued
follow up and search for pulsations may thus prove the companion is a
neutron star. 

Without a direct optical or radio detection at the moment, the nature of the
companion remains best investigated by comparing the \J\ system masses
to the collection of known \ac{DNS} and relativistic \ac{WD} binaries
with precise mass estimates (Table \ref{table:binaries}).  The
companion mass $m_c=\grmcompanion\,M_\odot$ is likely higher than that
of the most massive known similar \ac{WD}, the
$1.3^{+0.10}_{-0.10}\,M_\odot$ companion to PSR~B2303+46.  The
companion mass reported does,
however, fall well within the mass range of the recycled stars in
known \acp{DNS}.

\section{Implications and conclusion}
\label{sec:1906:implications}
We presented an updated timing solution for J1906+0746 that allows for
the measurement of three post-Keplerian parameters, $\dot{\omega}$,
$\gamma$ and $\dot{P}_b$.  
We measured pulsar and companion masses of
$m_p=$\,\grmpulsar$M_\odot$ and $m_c=$\,\grmcompanion$M_\odot$, respectively, 
compatible with a neutron-star or possibly a white-dwarf companion.

If the binary companion to this young unrecycled pulsar is a \ac{WD},
it must have formed first. Such systems are observationally
rarer by an order of magnitude than NS-WD systems in
which the NS is recycled. They also require a different mass
transfer history than the average binary pulsar.

The existence of \textit{young} pulsars in binaries around \acp{WD} was
predicted by \citet{dtws85} and
\citet{ty93}, and was subsequently confirmed by
the detections of the 12-day binary B2303+46 (\citealt{tamt93};
later identified as a NS-WD system by observation of the WD companion;
\citealt{vk99}) and the relativistic binary J1141$-$6545 \citep{klm+00,
  bbv08}, where the WD component was optically detected by \cite{abw+11}.  These systems do not fit in the traditional spin-up
scenario.

An evolutionary channel that \textit{can} explain this class is
outlined in \citet{ts00a}: the binary progenitor involves a primary
star with mass between $5$ and $11 M_\odot$ and a secondary with
initial mass between $3$ and $11M_\odot$.  The primary evolves and
overflows its Roche lobe, and the secondary accretes a substantial
amount of mass during this phase, which lasts $\sim$1 Myr.  At some
point after the primary forms a \ac{WD}, the now-massive secondary
evolves and a \ac{CE} is formed for a second, short, mass-transfer
phase.  The envelope is ejected, and a supernova occurs later, forming
the observed young pulsar. 

Pulsar \J\ and its possible WD companion would form a binary system of
widely different  binding energies.
In alternative theories of gravity,
particularly the Scalar-Tensor theories of gravity \citep{will93,
  de96}, such ``asymmetric'' systems emit dipolar
gravitational waves in addition to the quadrupolar emission predicted
by \ac{GR}. 
Such theories can thus be stringently tested by  measuring the orbital
decay in WD-NS systems \citep[as done for PSR~J1141$-$6545 by][]{bbv08}.
For symmetric systems like a double neutron star, which can have striking
precision in the measurement of their $\dot{P}_b$  \citep{ksm+06,wnt10},
only very little dipolar gravitational-wave 
emission is predicted by alternative theories of gravity.

Thus, for a pulsar-WD system, the closer the agreement of the
observed $\dot{P}_b$ with the value predicted by general relativity,
the stronger are the constraints on any alternative theories that
predict extra gravitational wave emission. Currently, the best limits
on such alternatives come from the measurement of the orbital decay of
the millisecond pulsar - WD system PSR~J1738+0333 \citep{fwe+12}. The main
limitation in the precision of this test 
is the precision (and accuracy) of the masses of the components of
that system, which was derived from optical spectroscopy and is
limited, to some extent, by uncertainties in the atmospheric models
used \citep{avk+12}. This is not so much a
limitation for PSR~J1906+0746, where the masses are known to very good
relative precision from the measurement of $\dot{\omega}$ and
$\gamma$. Therefore, if the companion of PSR~J1906+0746 is indeed a
WD, then this system could in principle provide a test of alternative
theories of gravity that is much superior to any current test.
However, for that to happen, three conditions must be fulfilled:
First, the companion should be confirmed to be a WD, as
\citet{abw+11} did for the companion of PSR~J1141$-$6545. {Second, the
parallax and proper motion should be precisely measured,
to better estimate the kinematic corrections to $\dot{P}_b$ (\S
\ref{pbdot_calc}).} Finally, the measurement of the orbital 
decay of the system should be improved via timing. The first condition
is hard to fulfill because {a WD companion is expected} to be very faint at
optical wavelengths (\S \ref{sec:1906:companion}). {The last two
conditions are hard to fulfill at the moment given the fact that the
pulsar is becoming very faint in the radio (\S \ref{sec:1906:HI}).}

If the companion is a recycled pulsar, there will not have been tidal
circularization of the current orbit; nor will there have been enough
time for gravitational-wave emission to significantly circularize it,
since the pulsar is only roughly $\tau_c = \characteristicage$\,kyrs
old (Table \ref{table:1906timing}).  Therefore the current
eccentricity $e = \eccentricity$, the lowest of any of the known
\acp{DNS}, must reflect the state of the orbit after the second
supernova.  
This implies a small supernova kick to the newborn young pulsar
\J, well within our upper limit on the system velocity.

For the general population of relativistic binaries there is a
selection effect favoring the detection of such low-eccentricity
systems: high eccentricities greatly increase the emission of
gravitational radiation, and those systems quickly coalesce after
their orbits have decayed \citep{cb05}. But for the detection of young
systems such as \J\ that selection effect has not yet developed. Only
binaries with eccentricities  $e > 0.94$ can expect to merge within
\J's age of $\tau_c = \characteristicage$\,kyrs
\citep{1964PhRv..136.1224P}.

The low eccentricity and system velocity, combined with the relatively
low mass of \J, suggest it was formed in an electron-capture, O-Ne-Mg
supernova \citep{2007AIPC..924..598V}.  In such a case, the spin axis
of the recycled pulsar is more likely to still be aligned with the
orbital angular momentum, in which case it will show little geodetic
precession (\S \ref{sec:1906:companion}). 
This formation scenario can thus be falsified by a future detection of
the recycled, companion pulsar if it \emph{does} precess into view.

In conclusion, we currently cannot confirm with certainty or rule out that the
companion of \J\ is a neutron star; and given the fast decline in
pulse flux due to geodetic precession, we will likely not improve on
our timing solution until the pulsar precesses back into view. Pulsar
\J\ is likely in a binary containing a double neutron star; or it is orbited
by a white dwarf, in a system formed by through an exotic binary interaction
involving two stages of mass transfer.

\section{Acknowledgments}

%% \emph{People:\\}
We thank
Jeroen Stil for advice on interpreting the VGPS data, 
Bryan Gaensler and Avinash Deshpande for discussions on the kinematic
absorption method, and Lindley Lentati for comparing timing algorithms.

The Arecibo Observatory is operated by SRI International under a
cooperative agreement with the National Science Foundation
(AST-1100968), and in alliance with Ana G. M\'endez-Universidad
Metropolitana, and the Universities Space Research Association. 

The National Radio Astronomy Observatory is a facility of the National
Science Foundation operated under cooperative agreement by Associated
Universities, Inc.

The Westerbork Synthesis Radio Telescope is operated by ASTRON with
support from the Netherlands Foundation for Scientific Research NWO.

The Nan\c cay radio Observatory is operated by the Paris Observatory,
associated to the French Centre National de la Recherche Scientifique
(CNRS). 

This work was supported by European Commission Grant
FP7-PEOPLE-2007-4-3-IRG-224838 (JvL) and by U.S. National Science
Foundation Grants AST-0647820 (DJN) and AST-0807556/AST-1312843 (JMW).
LK acknowledges the support of a doctoral Canada Graduate Scholarship.
Pulsar research at UBC is supported by an NSERC Discovery Grant and by
the Canada Foundation for Innovation.

\bibliographystyle{yahapj}

\bibliography{journals,psrrefs,modrefs,modjoeri,crossrefs,laura}

\end{document}